\documentclass[10pt]{article}
\usepackage{hyperref}
\hypersetup{colorlinks=true,linkcolor=black,citecolor=black,urlcolor=black,breaklinks=true}
\usepackage{float}
\usepackage[symbol]{footmisc}
\usepackage[superscript,sort]{cite}
\usepackage{array}
\usepackage{bm}
\usepackage{longtable}

\newcolumntype{x}[1]{%
	>{\centering\hspace{0pt}}p{#1}}%
\usepackage[normalem]{ulem} 
\usepackage{amsmath,amssymb}

\usepackage{amsthm,amscd,amsxtra,amsfonts,amsmath,amssymb,multirow}
\usepackage{wrapfig}
\usepackage[tiny,compact]{titlesec}
\usepackage{threeparttable} 
\usepackage{helvet}

\usepackage{booktabs}
\usepackage[table]{xcolor}
\usepackage{xcolor,colortbl}
\usepackage{placeins}
\usepackage{tikz}
\usetikzlibrary{shapes,arrows}
\usepackage[T1]{fontenc}
\usepackage[linesnumbered,ruled]{algorithm2e} 
\titlespacing*{\section}{0pt}{*0}{*0}
\titlespacing*{\subsection}{0pt}{*0}{*0}
\titlespacing*{\subsubsection}{0pt}{*0}{*0}
\titlespacing{\paragraph}{0pt}{*0}{*1}

\definecolor{MyPurple}{rgb}{1,0,1}

\newcommand{\beq}[1]{\begin{equation} \label{#1}}
\newcommand{\eeq}{\end{equation}}
\newcommand{\barray}{\begin{array}{ll}}
	\newcommand{\earray}{\end{array}}

\usepackage{longtable} 
\usepackage{threeparttable} 

\usepackage{array}
\newcolumntype{P}[1]{>{\centering\arraybackslash}p{#1}}

\usepackage{multirow}
\usepackage{color, colortbl}

\definecolor{Lightblue}{rgb}{0.867,0.914,0.961}
\definecolor{Lightgreen}{rgb}{0.883,0.934,0.848}

\DeclareGraphicsExtensions{.pdf,.jpeg,.png}

\title{Generative network complex for the automated generation of druglike molecules
}
 \author{Kaifu Gao$^{1}$, Duc Duy Nguyen$^{1}$, Meihua Tu$^{2}$, and  Guo-Wei Wei$^{1,3,4,}$\footnote{
		 Corresponding to Guo-Wei Wei.		Email: wei@math.msu.edu}\\
 $^1$ Department of Mathematics,
 Michigan State University, MI 48824, USA.\\
$^2$ Pfizer Medicine Design, 610 Main St, Cambridge, MA 02139, USA.\\
 $^3$ Department of Electrical and Computer Engineering,
 Michigan State University, MI 48824, USA. \\
 $^4$ Department of Biochemistry and Molecular Biology,
 Michigan State University, MI 48824, USA. \\
 }

\date{\today}

\begin{document}

\maketitle

\begin{abstract}
Current drug discovery is expensive and time-consuming. It remains a challenging task to create a wide variety of novel compounds with desirable pharmacological properties and cheaply available to low-income people.  In this work, we develop a generative network complex (GNC) to generate new drug-like molecules based on the multi-property optimization via the gradient descent in the latent space of an autoencoder. In our GNC, both multiple chemical properties and similarity scores are optimized to generate and predict drug-like molecules with desired chemical properties.   
To further validate the reliability of the predictions, these molecules are reevaluated and screened by independent 2D fingerprint-based predictors to come up with a few hundreds of new drug candidates. As a demonstration, we apply our GNC to generate a large number of new BACE1 inhibitors, as well as thousands of novel alternative drug candidates for eight existing market drugs, including Ceritinib, Ribociclib, Acalabrutinib, Idelalisib, Dabrafenib, Macimorelin, Enzalutamide, and Panobinostat. 
 
\end{abstract}

\section{Introduction}
Drug discovery ultimately tests our understanding of molecular biology, medicinal chemistry, genetics, physiology, and pharmacology, the status of biotechnology, the utility of computational sciences, and the maturity of mathematical biology. Technically, drug discovery involves target discovery, lead discovery, lead optimization, preclinical development, three phases of clinical trials, and finally, launching to the market only if a drug can be demonstrated to be safe and effective in every stage. Among them, lead discovery, lead optimization, and preclinical development disqualify tens of thousands of compounds based on their binding affinities, solubility, partition coefficients, clearances, permeability, toxicities, pharmacokinetics, etc., leaving only about tens of them to clinical trials. Therefore, drug discovery is expensive and time-consuming currently: it takes about \$2.6 billion dollars and more than ten years, on average, to bring a new drug to the market \cite{dimasi2016innovation}. Reducing the expense and speeding up the process is one of the top priorities of the pharmaceutical industry.

One of the key challenges in small molecule drug discovery is to find novel chemical compounds with desirable properties. Much effort has been taken to optimize this critical step in the drug discovery pipeline. For example, the development of high-throughput screening (HTS) has led to an unprecedented increase in the number of potential targets and leads \cite{hughes2011principles}. HTS can quickly conduct millions of tests to identify active compounds of interest using compound libraries rapidly \cite{macarron2011impact}. While there has been an increase in the number of potential targets and leads, the number of newly generated molecular entities has remained stable because of a high attrition rate by the elimination of leads with inappropriate physicochemical or pharmacological properties during preclinical development and clinical phases  \cite{graul2009overcoming,tareq2010predictions}. 
Rational drug design (RDD) approaches are proposed to better identify candidates with the highest probability of success \cite{waring2015analysis}. These methods aim at finding new medications based on the knowledge of biologically druggable targets \cite{dimasi2016innovation,stromgaard2017textbook}.

More recently, computer-aided drug design (CADD) has emerged as a useful approach in reducing the expense and period of drug discovery \cite{alqahtani2017silico}. Computational techniques have been developed for both virtual screening (VS) and optimizing the ADME properties of lead compounds. Primarily, these methods are designed as {\it in silico} filters to eliminate compounds with undesirable properties. These filters are widely applied for the assembly of compound libraries using combinatorial chemistry \cite{balakin2009compound}. The integration of early ADME profiling of lead chemicals has contributed to speeding up lead selection for phase-I trials without large amounts of revenue loss \cite{kapetanovic2008computer}. Currently, compounds are added in libraries based on target-focused design or diversity considerations \cite{huang2016protein}. VS and HTS can screen compound libraries to a subset of compounds whose properties are in agreement with various criteria \cite{szymanski2012adaptation}.

Despite these efforts, current databases of chemical compounds remain small when compared with the chemical space spanned by all possible energetically stable stoichiometric combinations of atoms and topologies in molecules. It is estimated that there are about 10$^{60}$ distinct molecules; among them, roughly 10$^{30}$ are drug-like \cite{macarron2011impact}. As a result, computational techniques are also being developed for {\it de novo} design of drug-like molecules \cite{schneider2005computer} and generating large virtual chemical libraries, which can be screened more efficiently for {\it in silico} drug discovery.

Among the available computational techniques, deep neural networks (DNN) have gathered much interest for their ability to extract features and learn physical principles from training data. Currently, DNN-based architectures have been successfully applied in a wide variety of fields in the biological and biomedical sciences \cite{min2017deep,mamoshina2016applications}.

More interestingly, many deep generative models based on sequence-to-sequence autoencoders (Seq2seq AEs) \cite{sutskever2014sequence}, variational autoencoders (VAEs) \cite{kingma2013auto}, adversarial autoencoders (AAEs) \cite{makhzani2015adversarial}, generative adversarial networks (GANs) \cite{goodfellow2014generative} or reinforcement learning \cite{kaelbling1996reinforcement}, etc. have been proposed for exploring the vast drug-like chemical space and generating new drug-like molecules.  Winter {\it et al.} \cite{winter2019efficient,winter2019learning} performed the optimization based on particle swarm optimization on the continuous latent space of a Seq2seq AE to generate new molecules with desired properties. Gomez-Bombarelli {\it et al.} \cite{gomez2018automatic} used a VAE to encode a molecule in the continuous latent space for exploring associated properties. Skalic {\it et al.} \cite{skalic2019shape} combined a conditional VAE and a captioning network to generate previously unseen compounds from input voxelized molecular representations. Kadurin {\it et al.} \cite{kadurin2017drugan} built an AAE to create new compounds. Sattarov {\it et al.} \cite{sattarov2019novo} combined deep autoencoder RNNs with generative topographic mapping to carry out {\it de novo} molecule design. A policy-based reinforcement learning approach was proposed to tune RNNs for episodic tasks \cite{olivecrona2017molecular,popova2018deep}, and extended to design desired molecules \cite{kang2018conditional}. Zhou {\it et al.} \cite{zhou2019optimization} also proposed a strategy to optimize molecules by combining domain knowledge of chemistry and state-of-the-art reinforcement learning techniques.

However, the generative strategies mentioned above are not drug-specified. What is vital to drug discovery is to design potential drug candidates for specific drug targets. In a regular drug discovery procedure, the starting point is target identification, followed by lead generation. Then, lead optimization is performed to make lead compounds more drug-like \cite{hughes2011principles}. 

It is useful to find new lead compounds to replace existing market drugs for several reasons. First, existing market drugs might not be optimal. For example,  they might not be potent enough, be too toxic and harmful to human health, or be too hard to synthesize. Additionally,  they might have side effects, insufficient aqueous solubility (log S) that reduces the absorption of drugs \cite{kerns2008vitro}, or higher partition coefficients (log P) that lead to improper drug distributions within the body  \cite{leo1971partition}. Moreover, for very expensive drugs, it is desirable to find cheap alternatives.

With the generation of new alternative lead compounds in mind, one can make use of existing drug datasets to develop drug-specified generative models. In this process, it is critical to apply  a similarity restraint to generate hundreds or even thousands of new drug-like molecules  inside the chemical space close to the reference molecule. This similarity restraint enables us to generate new molecules that remain effective to the target.  Moreover, from the viewpoint of machine learning, higher similarities to existing data always lead to more reliable predictions in generating molecules. The generative model can also realize lead optimization: by incorporating optimizers, generated molecules are designated to have one or more chemical properties better than the reference molecule. As a result, a large number of alternative drug candidates are created by drug-specified generative models. These candidates could be an effective and specified library to further screen for better or cheaper drug alternatives.

Therefore, in this work, we develop a generative network complex (GNC) based on the multiple-property optimization via gradient descent in the latent space to automatically generate new drug-like molecules. One workflow of our GNC consists of three following stages

\begin{enumerate}
	\item The SMILES string of a seed molecule are encoded into a vector in the latent space by a pre-trained encoder.
	\item Starting with the latent vector of the seed molecule, a DNN-based drug-like molecule generator modifies the vector via gradient descent, so that new latent vectors that satisfy multiple property restraints including chemical properties and similarity scores to the desired objectives are created.
	\item A pre-trained decoder decodes these new latent vectors into the SMILES strings of newly generated molecules.
\end{enumerate}

The rest of the paper is organized as follows. Section \ref{sec:methods} introduces our new GNC framework formed by the seq2seq AE and drug-like molecule generator. Section \ref{sec:experiments} discusses its reliability test on the BACE1 target, and more importantly, presents the performance of our GNC on a variety of existing drug targets. The insight into the roles of the multiple property restraints is offered in Section \ref{sec:discussions}. The conclusion is given in Section \ref{sec:conclusion}.

\section{Methods}\label{sec:methods}
\subsection{The sequence to sequence antoencoder (seq2seq AE)}
The seq2seq model is an autoencoder architecture originated from natural language processing  \cite{sutskever2014sequence}. It has been demonstrated as a breakthrough in language translation. The basic strategy of the seq2seq AE is to map an input sequence to a fixed-sized vector in the latent space using a gated recurrent unit (GRU) \cite{cho2014learning} or a long short-term memory (LSTM) network \cite{hochreiter1997long}, and then map the vector to a target sequence with another GRU or LSTM network. Thus the latent vector is an intermediate representation containing the ``meaning" of the input sequence.

In our case, input and output sequences are both SMILES strings -- a one-dimensional "language" of chemical structures \cite{weininger1988smiles}. The seq2seq AE is trained to have a high reconstruction ratio between inputs and outputs so that the latent vectors contain faithful information of the chemical structures (see the upper part of Figure \ref{fig:gen-seq2seq}). Here we utilize a pre-trained seq2seq model from a recent work \cite{winter2019learning}.

\subsection{The drug-like molecule generator based on the multi-property optimization}

\begin{figure}[h]
    \centering
	\includegraphics[width=0.7\textwidth]{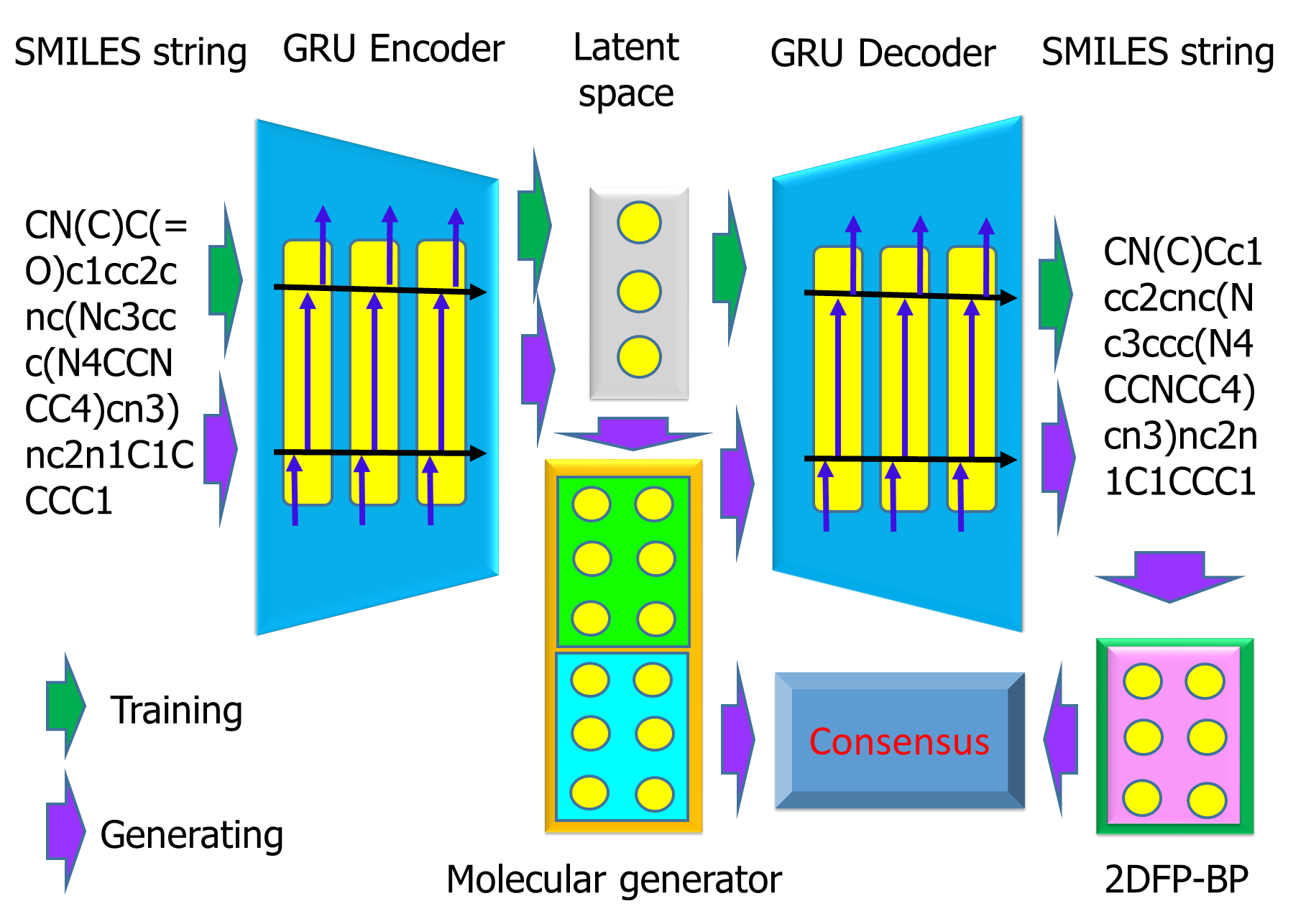}
	\caption{A schematic illustration of our generative network complex 2.}
	\label{fig:gen-seq2seq}
\end{figure}

In our new GNC, we design a drug-like molecule generator elaborately, so that generated molecules not only satisfy desired properties but also share common pharmacophores with reference compounds. From a seed molecule, one generative workflow of the GNC is depicted in Figure \ref{fig:gen-seq2seq} and described as below.

\begin{enumerate}
	\item Randomly pick a low-binding-affinity molecule from a target-specified training set as the seed, then the SMILES string of the seed molecule is encoded by a pre-trained encoder (in our case a GRU encoder) into a latent vector.
	\item The latent vector of the seed is fed into our DNN molecule generator. In every epoch, the generator comes up with a new vector $X \in \mathbb{R}^n$, and the deep learning network is instructed to evaluate $X$ via the following loss function
	
	\begin{equation}\label{eq:dnn_loss}
	{\cal L}(X)=\frac{1}{n}\sum_{i=1}^n k_i| \hat{y}_i(X) - y_{i0}|,
	\end{equation}
 
	where, $k_i$ is the $i$th predefined weight serving the purpose of emphasizing or deemphasizing different property restraints, $\hat{y}_i(X)$ is the predicted $i$th property value by a pre-trained predictor ${\cal M}_i$. Additionally, $y_{i0}$ is the objective value of the $i$th property. The restrained properties can be binding affinity (BA), the similarity score (Sim) to a reference molecule or others such as partition coefficient (Log P), Lipinski's rule of five \cite{lipinski1997experimental}, etc. Some guideline for $y_{i0}$ includes, in the BA restraint, one often sets $y_{\Delta G}$ < -9.6 kcal/mol, in the Log P condition, it is common to set $y_{\rm LogP}$ < 5, etc.
	
	\item Gradient descent is used to minimize the loss function in Eq. (\ref{eq:dnn_loss}) until the maximum number of epochs is reached.
	
	\item The generated latent vectors satisfying the desired restraints are decoded into SMILES strings through a pre-trained decoder, as shown in Figure \ref{fig:gen-seq2seq}. 

\end{enumerate}

\begin{figure}[h]
	\centering
	\includegraphics[width=0.7\textwidth]{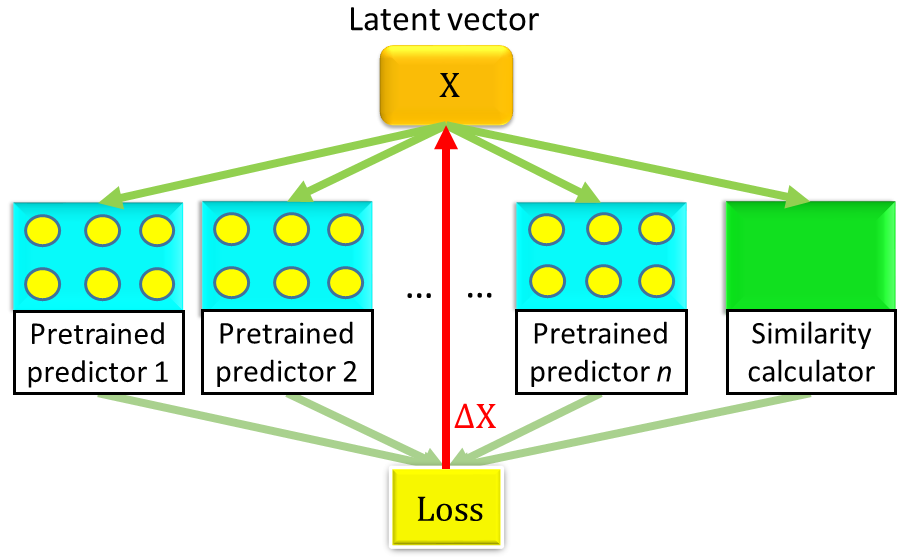}
	\caption{The illustration of the latent-space molecule generator.}
	\label{fig:gen-mechanism}
\end{figure}

To create a variety of novel drug-like molecules originated from leads or existing drugs (reference molecules), one can adopt different seed molecules as well as different objective values to achieve desired properties and similarity scores. The ultimate purpose of our molecule generator is to keep modifying the latent vector to satisfy the multiple druggable property restraints. Figure \ref{fig:gen-mechanism} illustrates the general mechanism of our generator. Notably, the pre-trained predictor weights inside our model stay intact throughout the backpropagation of the training process to the generator. The loss function converges when all conditions are met, and then, resulting latent vectors are decoded into SMILES strings.

\subsection{The parameters of the molecule generator} \label{sec:param-gen}
In our model, the dimension of the latent space is 512, so the input and output dimensions of the DNN molecule generator are also 512. The DNN generator has two hidden layers, with 1024 neurons in each layer. The activation function is tanh, the learning rate is 0.1, and the momentum is also 0.1. In this work, we are interested in binding affinity and similarity score restraints. The regularization coefficients of these two restraints ($k_{\rm \Delta G}$ and $k_{\rm Sim}$) are set to 1 and 10, respectively. The similarity score restraints is determined via the Tanimoto coefficient between a generated latent vector and the latent vector of a reference molecule. 

The binding affinity restraints rely on pre-trained binding affinity predictors. A pre-trained binding affinity predictor (LV-BP) takes latent vectors as its inputs and return predicted binding affinities. Therefore, typically, the input dimension of the predictor is 512, the output dimension is 1. The DNN predictor has three hidden layers with 1024, 1536, and 1024 neurons, respectively. The ReLU activation function is applied. The learning rate is 0.001, the number of training epochs is 4000, the batch size is 4. The predictor network is trained on target-specified datasets carefully selected from public databases such as ChEMBL \cite{gaulton2011chembl}. The generator and predictor are both programmed in the framework of PyTorch (Version 1.0.0) \cite{paszke2017pytorch}. 

In the current work, for each generation task, we randomly pick 50 low-binding-affinity molecules from the preselected dataset as seeds. For each seed, the generative network is run in a total of 2000 epochs, which takes less than 10 minutes under the supercomputer equipped with one Nvidia K80 GPU card. In practice, to quickly fill up the potential chemical search space, one can use more seeds and run more epochs for each seed.

\subsection{Binding affinity reevaluation by the 2D-fingerprint predictor} \label{sec:2DFP-BP}
Besides generating new molecules, the LV-BP in our GNC also predicts their binding affinities. However, no experimental values are available to validate these predicted affinities. Therefore, we cross-validate them using alternative binding affinity predictors. In the present work, we construct machine learning predictors based on 2D fingerprints (2DFP-BPs) to reevaluate the affinities of generated compounds. The 2D fingerprints computed from their SMILES strings are inputs to these 2DFP-BPs. If the predictions from the LV-BP and 2DFP-BPs are consistent, we regard the predictions as reliable.

According to our previous tests \cite{gao20202d}, the consensus of ECFP4 \cite{rogers2010extended}, Estate1\cite{hall1995electrotopological} and Estate2 \cite{hall1995electrotopological} fingerprints performs best on binding-affinity prediction tasks. Therefore, this work also makes use of this consensus. We employ the RDKit software (version 2018.09.3) \cite{landrum2006rdkit} to generate 2D fingerprints from SMILES strings.  Since the training sets in our current cases are not so large, we apply gradient boosting decision tree (GBDT) \cite{schapire2003boosting} model due to its accuracy and speed when handling small datasets. This GBDT predictor is constructed using the gradient boosting regressor module in scikit-learn (version 0.20.1) \cite{pedregosa2011scikit} and the following parameters: n\_estimators=10000, max\_depth=7, min\_samples\_split=3, learning\_rate=0.01, subsample=0.3, and max\_features=sqrt.

The criteria used in our reevaluation are the root mean square error (RMSE), Pearson correlation coefficient ($R_p$), and active ratio. Here, the active ratio means the ratio of the number of the active molecules indicated both by the 2DFP-BP and LV-BP to the number of the active ones indicated by the LV-BP.

\subsection{Multitask DNN predictors} \label{sec:param-multi}

Multitask DNN predictors \cite{caruana1997multitask} for both latent vectors and 2D fingerprints are built for the drug Ribociclib with two different targets. 

The latent-vector based model has three hidden layers with 1024, 1536, and 1024 neurons. For the 2D-fingerprint based models, because the three different 2D fingerprints ECFP4, Estate1, Estate2 have 2048, 79, and 79 features, respectively, two different network architectures are used. For ECFP4, the numbers of neurons in the three hidden layers are 2500, 1500, and 500, respectively. For Estate1 and Estate2, their numbers of neurons are 500, 1000, and 500. Other parameters are the same as those of our single task predictors. These multitask models are also programmed in the framework of PyTorch (Version 1.0.0).

\subsection{Drug-target interaction and common pharmacophore analysis}

The interactions between drugs and their targets, as well as the pharmacophores of the drugs, are investigated. The purpose is to explore whether our generated molecules can still bind to the drug targets.

Drug-target interactions are analyzed via the protein-ligand interaction profiles \cite{salentin2015plip}. It can identify drug-target interactions as well as their types such as hydrogen bonds, hydrophobic interactions, etc.

However, the interaction analysis itself could not determine whether interactions are critical or not to the drug-target binding. By using the Phase module in Schr\"{o}dinger (version 2018-4) \cite{dixon2006phase}, we build pharmacophore models via searching common pharmacophores in all the active compounds to the target. Since these pharmacophores are widespread to all the active compounds, they are critical to the binding. Thus, if generated molecules still contain such pharmacophores, we can believe they are potential binders.

It could be time-consuming to recognize common pharmacophores of hundreds of compounds. To avoid this obstacle, we group compounds into 50 clusters via the k-means algorithm implemented by scikit-learn \cite{pedregosa2011scikit}. We, then, collect the centroids of these 50 clusters for the common pharmacophore search.

\subsection{Datasets}
In this work, first, we explore the effect of different objective values in our generator on the binding-affinity prediction reliability to generated molecules. We carry out this reliability test on the Beta-Secretase 1 (BACE1) dataset.  

BACE1 is a transmembrane aspartic-acid protease human protein encoded by the BACE1 gene. It is essential for the generation of beta-amyloid peptide in neural tissue \cite{vassar2009beta}, a component of amyloid plaques widely believed to be critical in the development of Alzheimer's, rendering BACE1 an attractive therapeutic target for this devastating disease \cite{prati2017bace}. We download 3916 BACE1 compounds from the ChEMBL database. In the seq2seq autoencoder we utilized, there is a molecule filter that only selects organic molecules with more than 3 heavy atoms, their weights between 12 and 600, and their Log P values between -5 and 7 \cite{winter2019learning}. As a result, a total of 3151 molecules in the BACE1 dataset pass this filter and are used as the training set. 

\begin{table}[h]
	\centering
	\begin{tabular}{|P{2.0cm}|P{1.5cm}|P{3.2cm}|P{1.4cm}|P{1.4cm}|P{2.0cm}|P{2.6cm}|}
		\hline
		\rowcolor{Lightgreen}
		\textbf{Drug name} & \textbf{ChEMBL ID} & \textbf{Treatment} & \textbf{FDA approval year} & \textbf{$\Delta G$ (kcal/mol)} & \textbf{Filtered training set size} & \textbf{$\Delta G$ range of training set (kcal/mol)} \\
		\hline
		Ceritinib & 2403108 & Non-small cell lung cancer & 2014 & -10.77 & 1203 & -5.26 to -13.93 \\ \hline
		Ribociclib & 3545110 & Breast cancer & 2017 & -10.98 \& -10.16 & 918 \& 289 & -6.31 to -12.98 \& -4.11 to -14.13  \\
		\hline
		Acalabrutinib & 3707348 & Mantle cell lymphoma & 2017 & -11.67 & 1451 & -4.21 to -14.05  \\ \hline
		Idelalisib & 2216870 & Blood cancers & 2014 & -10.43 & 1959 & -1.92 to -14.16  \\ \hline
		Macimorelin & 278623 & Adult growth hormone deficiency & 2017 & -10.48 & 608 & -4.18 to -14.68  \\ \hline
		Dabrafenib & 2028663 & Cancers with a mutated gene BRAF & 2013 & -12.35 & 2254 & -5.49 to -14.68  \\ \hline
		Enzalutamide & 1082407 & Prostate cancer & 2012 & -10.53 & 1386 & -3.83 to -13.72  \\ \hline
		Panobinostat & 483254 & Cancers & 2015 & -12.46 & 1645 & -2.91 to -12.46  \\
		\hline
	\end{tabular}
	\caption{The information about the eight market drugs used in the present study.}
	\label{table:eight-drug-inf}
\end{table}

More importantly, we employ our GNC to design alternative promising drug candidates for the eight drugs on the market. For each drug, we construct a dataset of the compounds that bind to the same drug target from the ChEMBL database. The collected compounds are also filtered by the filter in the seq2seq autoencoder. Table \ref{table:eight-drug-inf} lists information regarding these eight drugs, namely drug name, ChEMBL ID, FDA approval year, experimental drug affinity ($\Delta G$), filtered training set size, and affinity range of the training set.

The SciFinder is one of the most comprehensive databases for chemical literature and substances (\url{ https://scifinder-n.cas.org}). It contains more than 143 million organic and inorganic substances. Here we search our generated molecules in this database to confirm they are new and never existed before.

\section{Experiments} \label{sec:experiments}
\subsection{Designing BACE1 inhibitors}
\subsubsection{The accuracy of the seq2seq AE and predictors}
We first test the accuracy of the seq2seq autoencoder and the LV-BP and 2DFP-BP predictors.

When performing the seq2seq model on the filtered BACE1 dataset with 3151 molecules, the reconstruction ratio is 96.2\%. This high ratio guarantees the essential information of these input molecules is encoded into the corresponding latent vectors. 

Subsequently, these latent vectors are used as the features to train our latent-vector DNN binding affinity predictor (LV-BP), the labels are their corresponding experimental binding affinities. In a 5-fold cross-validation test on the BACE1 dataset, the LV-BP achieves an average Pearson correlation coefficient ($R_p$) of 0.871 and an average RMSE of 0.704 kcal/mol.

The 2D-fingerprint GBDT binding affinity predictor (2DFP-BP) is used to reevaluate the predictions from the LV-BP. We also test this 2DFP-BP by the 5-fold cross-validation. The average $R_p$ and RMSE are 0.874 and 0.692 kcal/mol, respectively, quite comparable to the LV-BP.

\subsubsection{Convergence analysis}

\begin{figure}[!htp]
	\centering
	\includegraphics[width=0.6\textwidth]{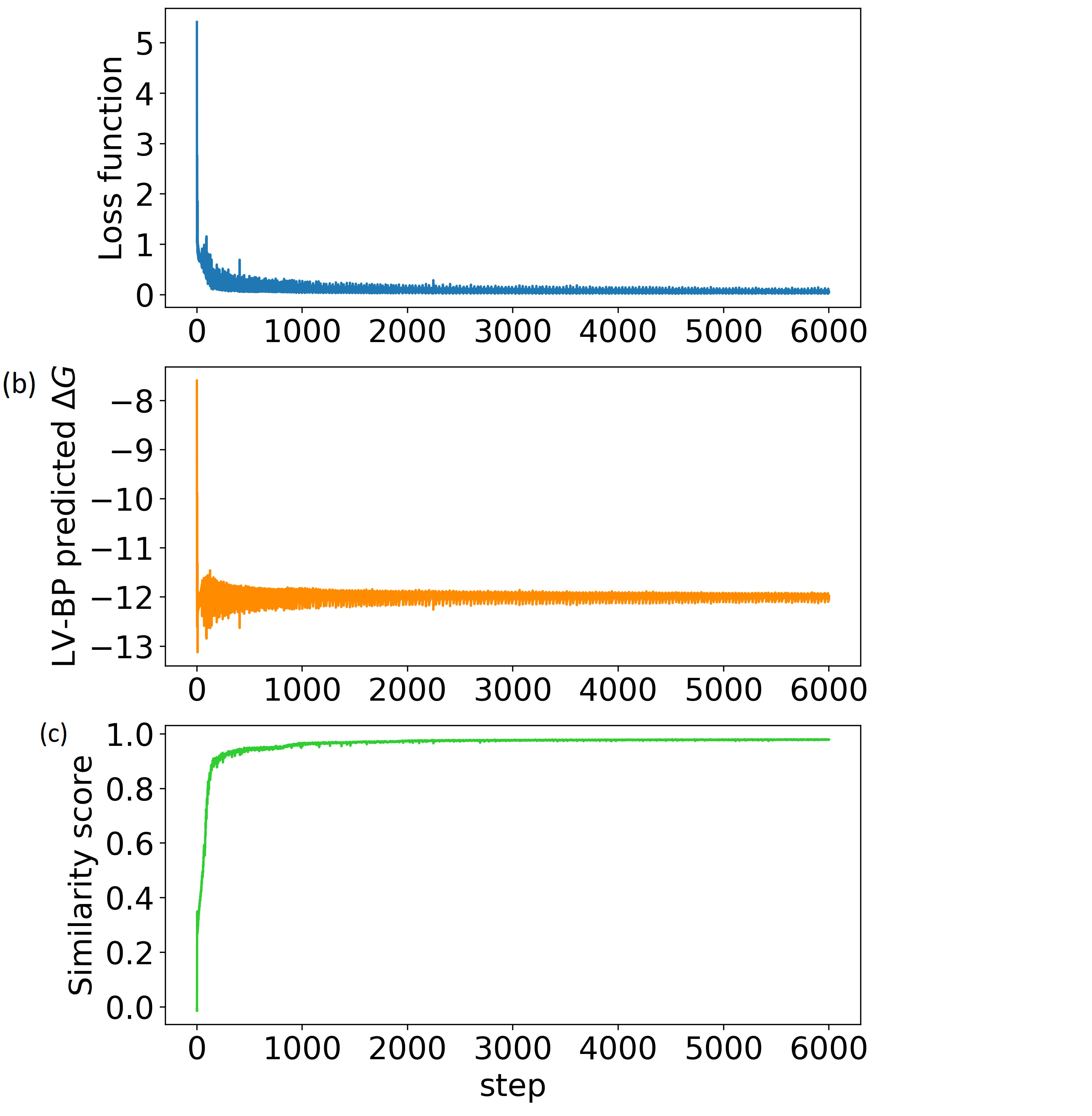}
	\caption{The convergence of the loss function, the LV-BP predicted $\Delta G$s, and the similarity score to the reference molecule during a molecule generation course. In this example, the $\Delta G$s of the seed molecule and the reference molecule are -6.81 kcal/mol and -12.02 kcal/mol, respectively. To force generated molecules evolving towards the reference molecule, we set the similarity score objective and the $\Delta G$ objective to be 1.00 and -12.02 kcal/mol, respectively.}
	\label{fig:pathway-loss}
\end{figure}

\begin{figure}[!htp]
	\centering
	\includegraphics[width=0.8\textwidth]{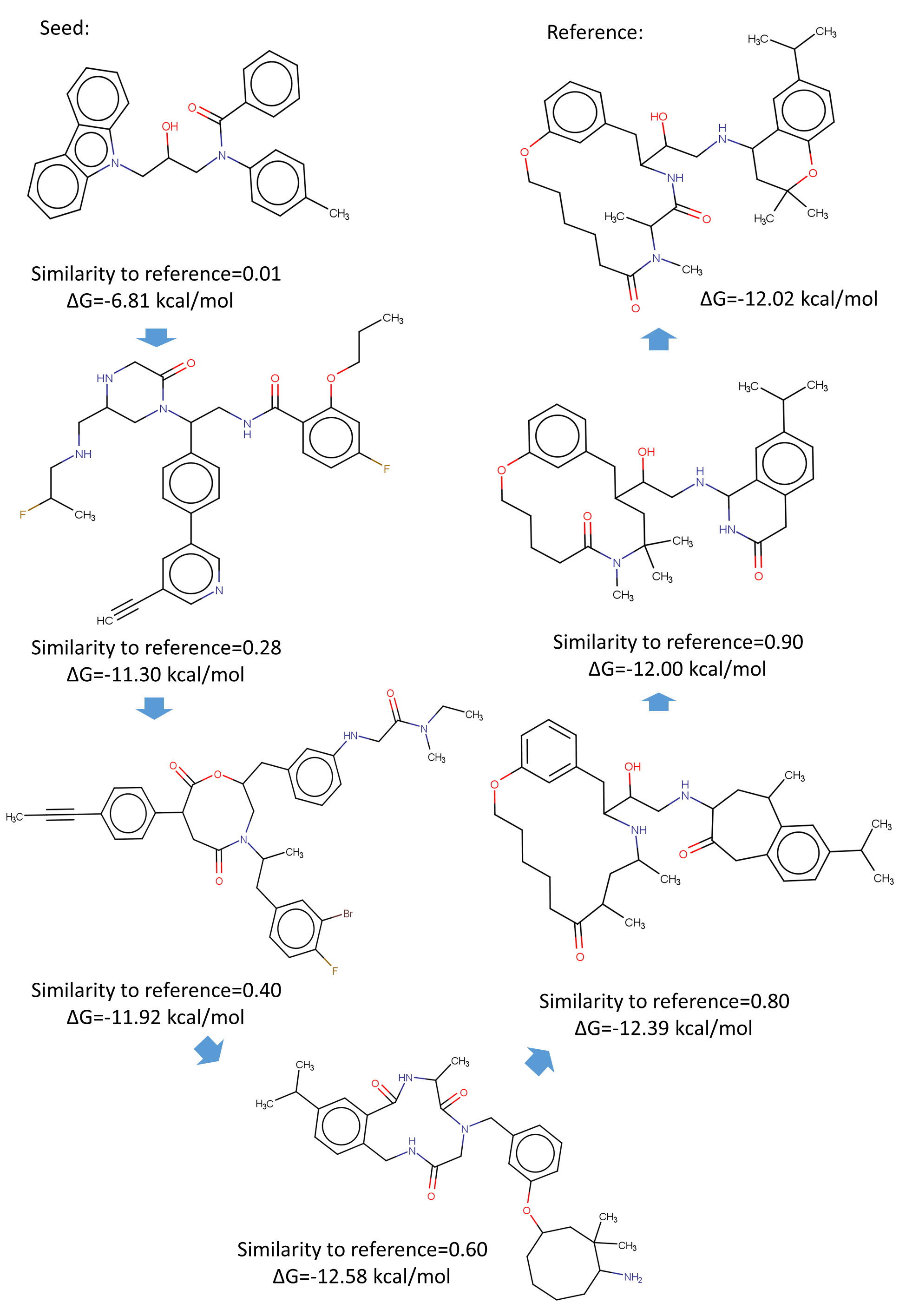}
	\caption{A series of generated molecules over the evolution from the seed to the reference molecule.}
	\label{fig:gen-pathway}
\end{figure}

Here we conduct our GNC to generate new molecules and analyze how these new molecules evolve over a generation course. We start this experiment with a seed molecule picked from the BACE1 dataset; this molecule is far from active with the binding free energy ($\Delta G$) = -6.81 kcal/mol. The reference molecule is also from the BACE1 dataset, it is highly active with $\Delta G$ = -12.02 kcal/mol. The binding affinity objective $y_{\Delta G}$ is set to be -12.02 kcal/mol, and the similarity score to the reference molecule is targeted to $y_{\rm sim}=1.0$. 

Figure \ref{fig:pathway-loss}a depicts the loss function values computed at every epoch; Figure \ref{fig:pathway-loss}b and c illustrate the LV-BP predicted binding affinities and similarity scores, respectively. From these figures, one can observe that our GNC produces new potential BACE1 inhibitors with desirable binding affinities in less than 3000 epochs.

Figure \ref{fig:gen-pathway} shows a series of generated molecules over the evolution from the seed to the reference molecule. The starting point is the seed molecule, its binding affinity and similarity score to the reference molecule are as low as -6.81 kcal/mol and 0.01, respectively. By receiving the feedback from the gradient descent in the generator, the similarity score gradually rises to 1.0. The improvement to the binding affinity is even faster: while a created molecule has a similarity score of 0.28, its LV-BP predicted $\Delta G$ already reaches -11.30 kcal/mol; while the similarity score is 0.90, the LV-BP predicted $\Delta G$ is -12.00 kcal/mol, which is essentially the same as the reference molecule's $\Delta G$ of -12.02 kcal/mol.

\subsubsection{Reliability test on the designed BACE1 inhibitors}\label{sec:BACE1-reliablity-test}
\begin{figure}[!htb]
	\centering
	\includegraphics[width=0.4\textwidth]{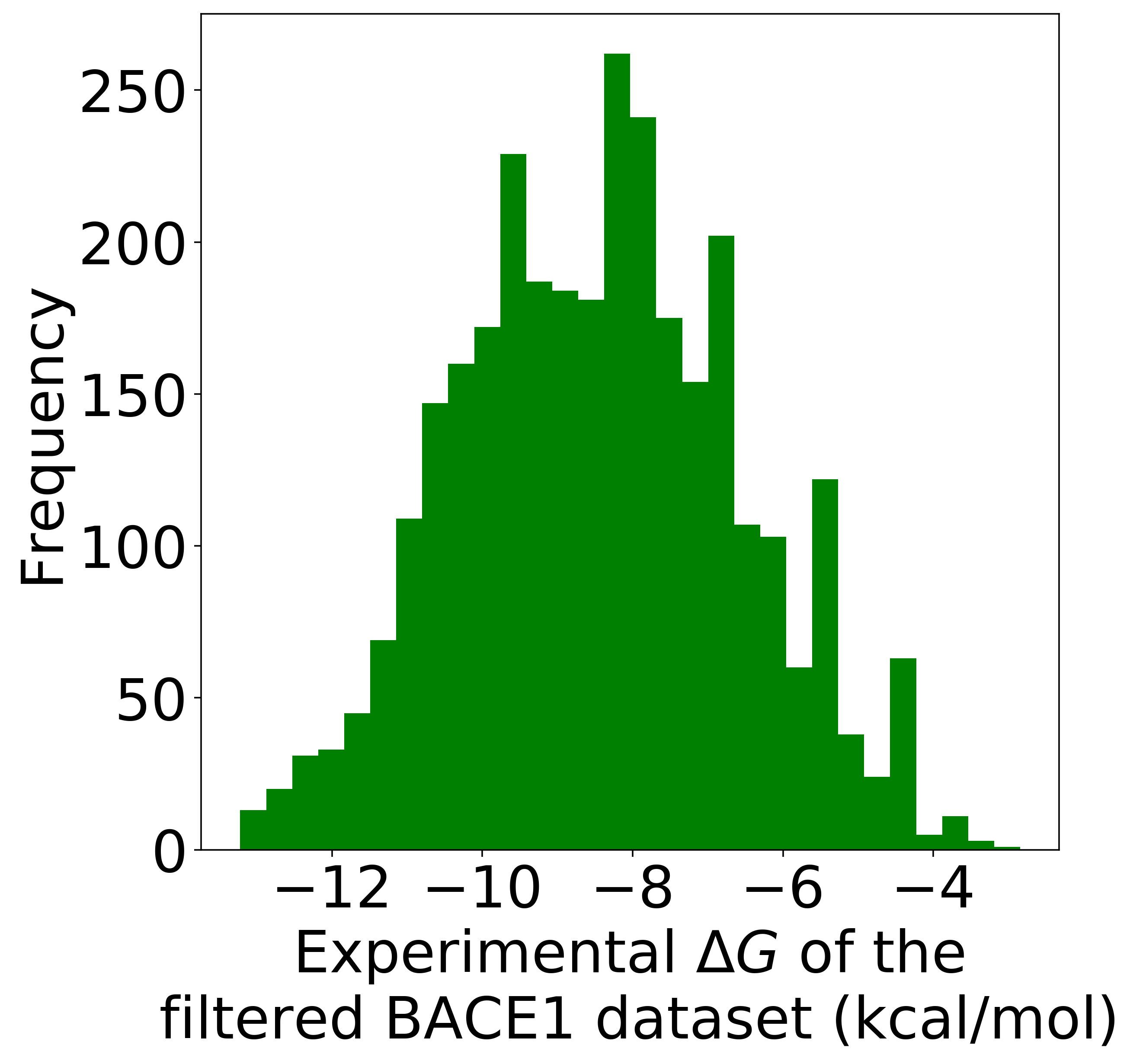}
	\caption{The experimental $\Delta G$ distribution of the filtered BACE1 dataset.}
	\label{fig:bace_dist}
\end{figure}

Using our GNC, we also generate millions of compounds targeting a wide range of binding affinities and similarity scores. Then the prediction reliability with these different ranges of binding affinities and similarity scores is tested. Individually, the similarity score objectives, $y_{\rm sim}$, vary from 0.50 to 0.95 with an increment of 0.025; the binding affinity objectives, $y_{\Delta G}$, receive values from -9.6 kcal/mol to -13.1 kcal/mol with an increment of -0.25 kcal/mol. Here we select -9.6 kcal/mol as the starting point since this value is a widely accepted threshold to identify active compounds; the endpoint of $\Delta G$ = -13.1 kcal/mol is the highest binding affinity value in the BACE1 dataset (see Figure \ref{fig:bace_dist}).

\begin{figure}[!htb]
	\centering
	\includegraphics[width=1.0\textwidth]{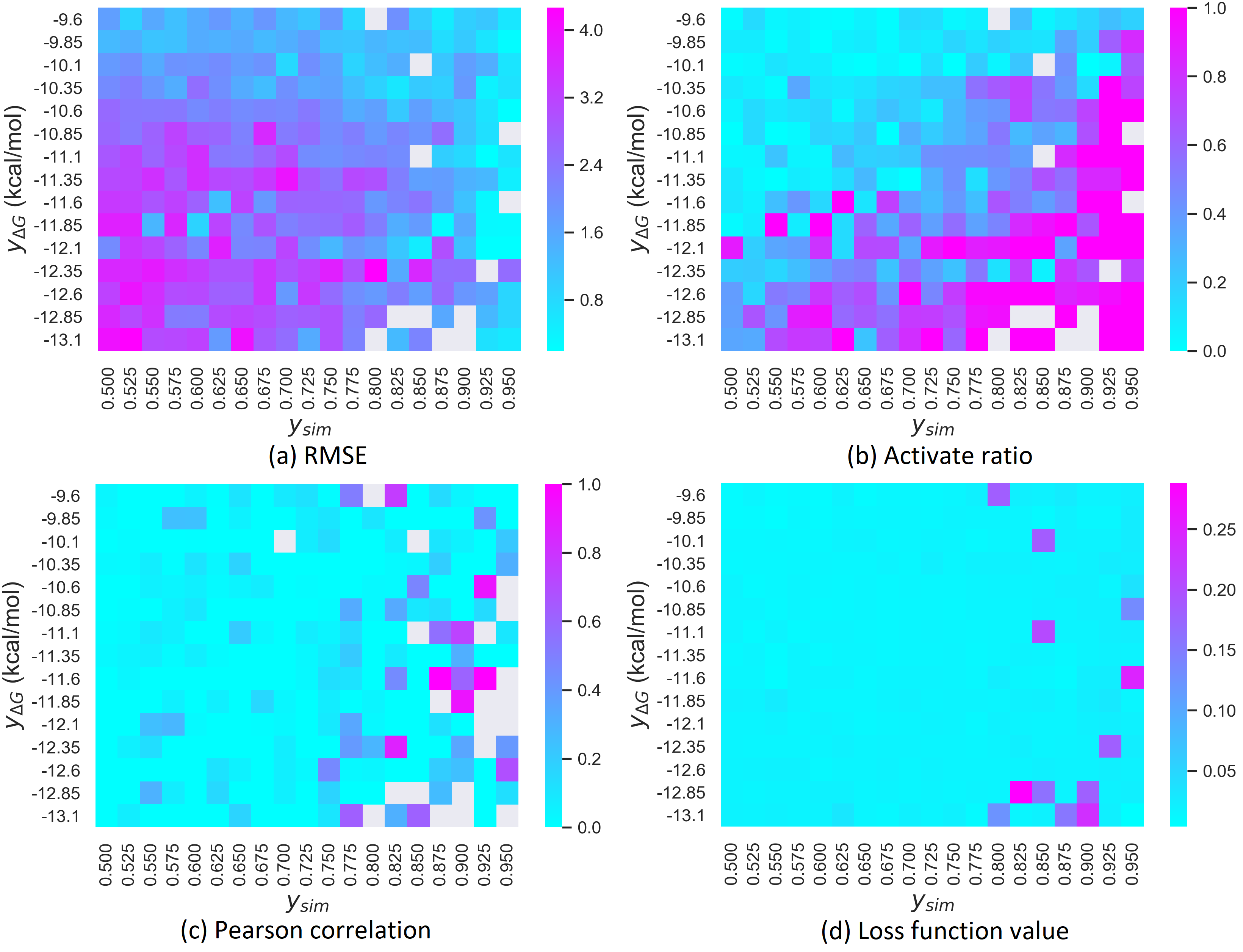}
	\caption{The reliability test to our GNC generator on the BACE1 dataset. The prediction reliability with different binding affinity objectives ($y_{\Delta G}$) and similarity score objectives ($y_{\rm sim}$) is tested. The discrepancies between the LV-BP and the 2DFP-BP predictions are evaluated by different criteria: (a) The RMSE; (b) The activate ratio; (c) The Pearson correlation; (d) The loss function values.}
	\label{fig:bace-reeval}
\end{figure}

The reliability test is based on reevaluating the LV-BP predicted binding affinities by the 2DFP-BP. The reliability criteria are the RMSE, active ratio, and $R_p$ between the LV-BP and 2DFP-BP prediction. The heatmaps in Figure \ref{fig:bace-reeval} shows these evaluation metric values corresponding to different similarity score and binding affinity restraints. Few blank points are present in each heatmap due to no available generation meeting these specific restraints.

Figure \ref{fig:bace-reeval}a plots the RMSE metrics. It reveals the most reliable region, i.e., having low RMSE, is $y_{\rm sim}$ above 0.925, and the $y_{\Delta G}$ between -10.1 kcal/mol and -12.1 kcal/mol. This is expectable since machine learning models can render accurate predictions if generated structures are highly similar to the training data (see Figure \ref{fig:bace_dist}). Besides, a large population of training samples have $\Delta G$s between -10.1 kcal/mol and -12.1 kcal/mol, leading more reliable predictions to molecules generated inside this range. Outside this range, as both $y_{\Delta G}$ and $y_{\rm sim}$ decreasing, the RMSEs between the LV-BP and 2DFP-BP predictions increase. Specifically, if $y_{\Delta G} < -12.1$ kcal/mol and $y_{\rm sim} < 0.675$, the RMSEs are always over 3.2 kcal/mol. 

Figure \ref{fig:bace-reeval}c depicts the $R_p$s between the LV-BP and 2DFP-BP predictions with respect to $y_{\Delta G}$ and $y_{\rm sim}$. Similar to the manner of the RMSE distribution, $-$12.1   kcal/mol $ \leq y_{\Delta G} \leq 10.1$ kcal/mol and $0.9\leq y_{\rm sim} \leq 0.95$ lead to the $R_p$ values consistently higher than 0.8.

The last component of our reliable analysis regards the loss function magnitudes of our GNC generator plotted in Figure \ref{fig:bace-reeval}d. In most cases, the loss function values are less than 0.05. However, in some special situations, our network cannot maintain the loss values lower than 0.15. At these points, we cannot find any generated molecules subject to the multi-property restraints simultaneously, which renders the blank spots in the criteria plots in Figures \ref{fig:bace-reeval}a, b, and c.

In summary, to generate molecules with reliable predictions, one should set the binding affinity objective $y_{\Delta G}$ in a region filled with a large population of training data. Besides, the similarity score restraint $y_{\rm sim}$ should be high. However, in some circumstances, the generated molecules that have high predicted affinities should also be included in further consideration.

\subsection{Designing alternative drug candidates}

In this section, we utilize our GNC to produce alternative drug-like molecules with high binding affinities to the existing drugs' targets, which provides effective libraries for further improvement or searching cheaper drug alternatives. This work discusses eight drugs and their targets with the information regarding the names, ChEMBL IDs, energies, etc. summarized in Table \ref{table:eight-drug-inf}. All of these drugs were approved by the FDA in the recent decade to treat critical diseases, especially a variety of cancers. Notably, the drug Ribociclib has two different targets (Cyclin-dependent kinase 4 and Cyclin-dependent kinase 6), so Ribociclib has two sets of $\Delta G$s and two sets of training compounds.

\subsubsection{The single-target drug: Ceritinib}
\paragraph{The statistics of the drug Ceritinib.} The brand name of Ceritinib (ChEMBL ID: CHEMBL2403108) is Zykadia. It was developed by Novartis and approved by the FDA in April 2014 to treat different types of non-small cell lung cancers. Ceritinib is extremely expensive, with the monthly cost of Ceritinib-based treatment in the US being approximately \$11,428.

\begin{figure}[!htb]
	\centering
	\includegraphics[width=0.7\textwidth]{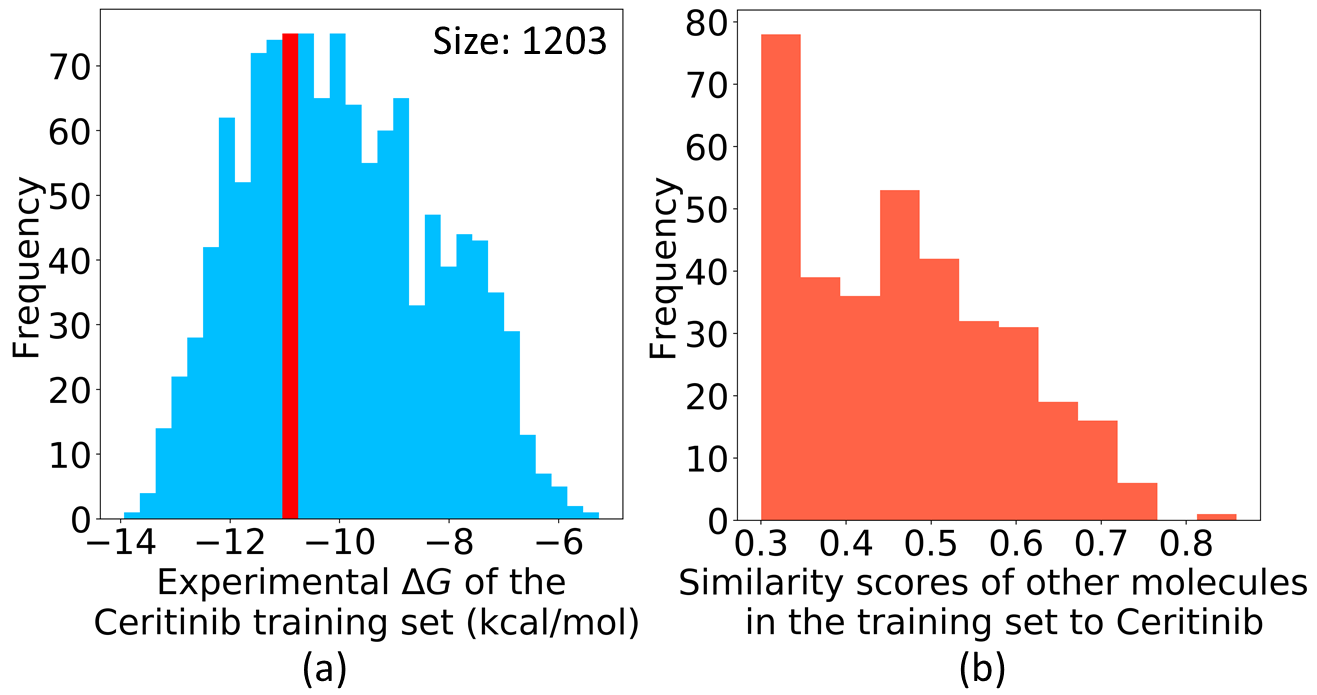}
	\caption{(a) The $\Delta G$ distribution of the filtered training set to the target ALK. The red bar indicates the interval containing the $\Delta G$ of Ceritinib. (b) The similarity scores of the other molecules in this set to Ceritinib.}
	\label{fig:chembl2403108-training-dis}
\end{figure}

Ceritinib inhibits the ALK tyrosine kinase receptor (ALK, CHEMBL ID: CHEMBL4247). The ChEMBL database provides 1407 molecules with experimental binding affinity labels to this target available. After going through the filter in our model, 1203 molecules are left to train our generator and predictor. Figure \ref{fig:chembl2403108-training-dis}a depicts the $\Delta G$ distribution of this training set; it unveils that, hundreds of training samples have their binding affinities close to or even higher than Ceritinib. The similarity score distribution in Figure \ref{fig:chembl2403108-training-dis}b indicates, the training set includes 355 samples with their similarity scores to Ceritinib over 0.3, 56 samples with such scores over 0.6. These promising analytical assessments enable our GNC to design potential inhibitors to the target ALK.

\paragraph{Designing new druglike molecules.}
Here we use Ceritinib as the reference molecule to design alternative Ceritinib drugs. Section \ref{sec:BACE1-reliablity-test} suggests, to generate new molecules with desirable properties, the binding affinity objectives should be inside a region with plenty of training data, and the similarity scores to the reference molecule $y_{\rm sim}$ should be restrained to be high. However, high similarity scores could lead to quite limited chemical space. Our solution to this drawback is, first, extend similarity score restraint to a broader range, then reevaluate generated compounds using the 2DFP-BP, and only pick the ones with low discrepancies between the LV-BP and 2DFP-BP predicted affinities.

Following this strategy, we set the similarity score restraints varying from 0.3 to 0.9 with an increment of 0.025; this is because more than 300 training samples have their similarity scores over 0.3, which is supportive to generate compounds in this similarity range. The binding affinities are aimed at the interval from -10.5 kcal/mol to -12.25 kcal/mol with an increment of -0.25 kcal/mol; this binding affinity region covers hundreds of training samples as well as the drug Ceritinib itself. After reevaluating by the 2DFP-BP, any generated molecules with the relative errors between the LV-BP and 2DFP-BP predicted affinities over 5 \% are thrown away.

\begin{figure}[h]
	\centering
	\includegraphics[width=0.55\textwidth]{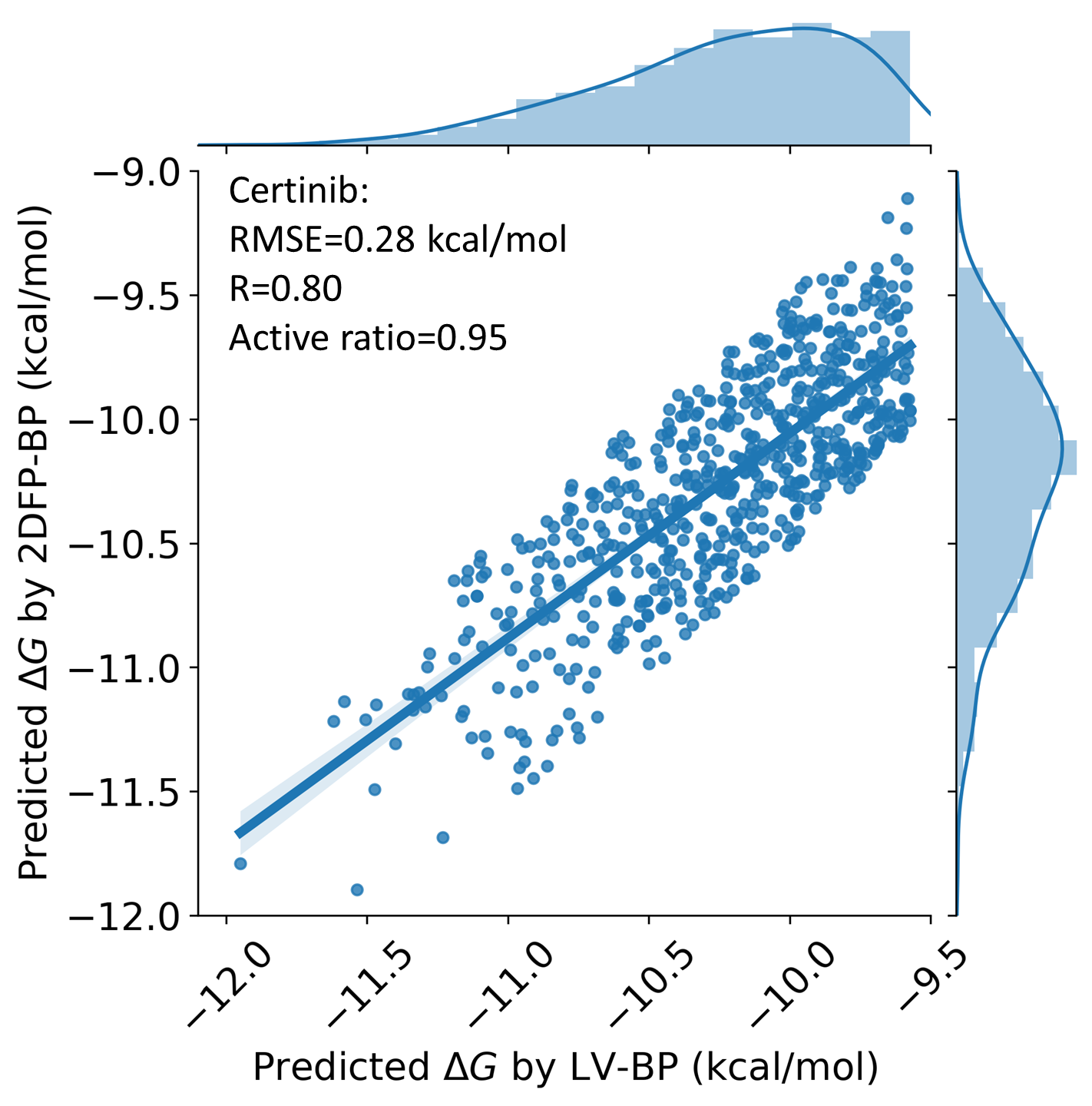}
	\caption{The correlation plot between the LV-BP and 2DFP-BP predicted $\Delta G$s of the generated molecules to the target ALK, the ones having high relative errors between the two predictions (>5 \%) are already eliminated.}
	\label{fig:corr-2d-dl-2403108}
\end{figure}

\paragraph{The 2DFP-BP reevaluation.} The details of the 2DFP-BP are offered in Section \ref{sec:2DFP-BP}. With Ceritinib as the reference, our GNC model creates 1095 novel active drug-like molecules, upon eliminating the ones with high discrepancies in their LV-BP and 2DFP-BP predictions, 629 molecules are left. Figure \ref{fig:corr-2d-dl-2403108} indicates, for these 629 molecules, the correlation between the two predictions is quite promising, with the RMSE = 0.28 kcal/mol, $R_p$ = 0.80, and active ratio = 0.95, respectively. This statistical information endorses the drug-likeness potential of our AI-generated molecules.

\begin{figure}[!htb]
	\centering
	\includegraphics[width=0.8\textwidth]{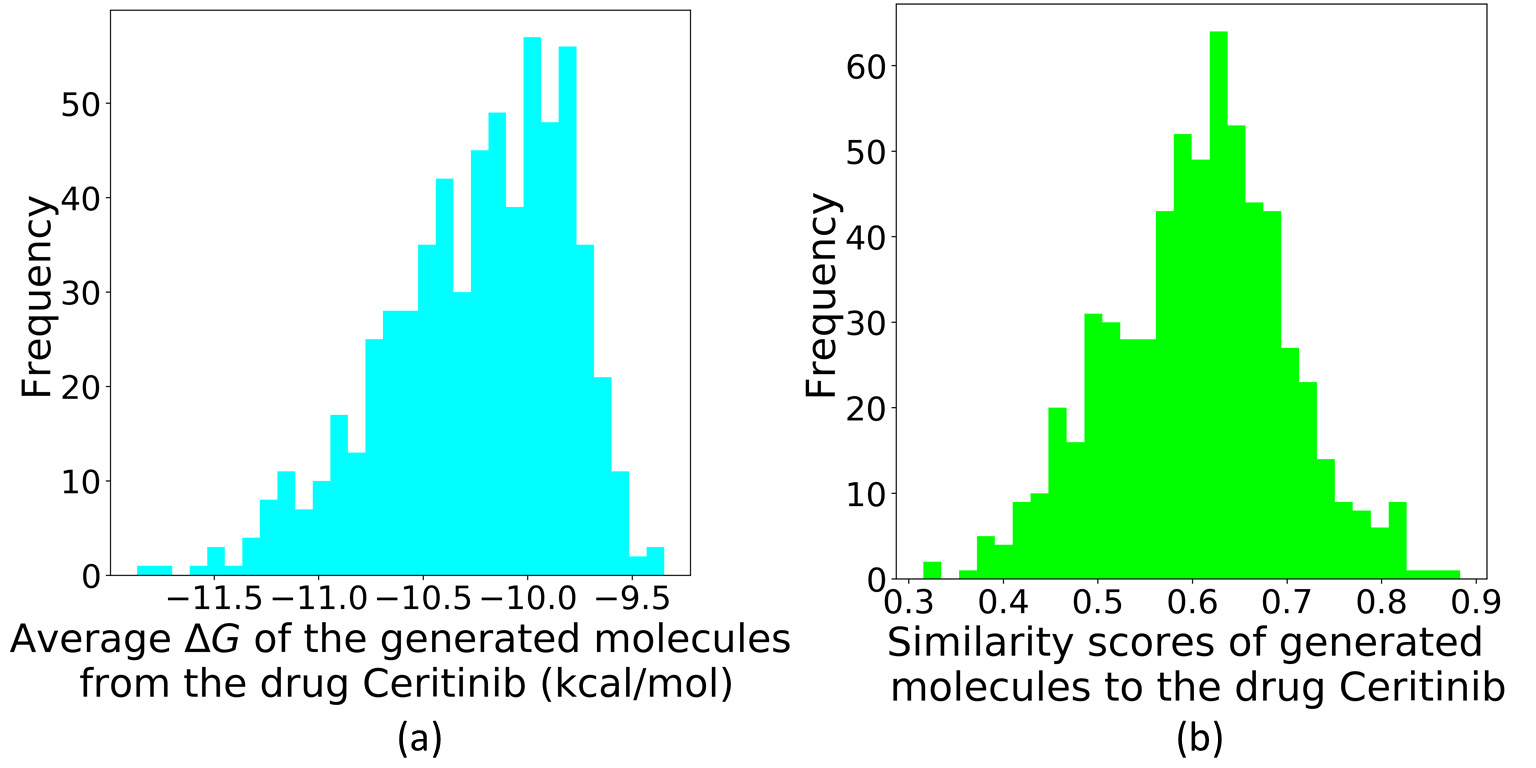}
	\caption{The average predicted binding affinities to the target ALK of the generated molecules and their similarity scores to the reference drug Ceritinib. (a) The distribution of the averages of the LV-BP and 2DFP-BP predicted $\Delta G$s to the target ALK. (b) The similarity score distribution to the reference drug Ceritinib.}
	\label{fig:chembl2403108-gen-dis}
\end{figure}

The LV-BP and 2DFP-BP predicted binding affinities of these 629 molecules are also averaged, and their distribution is shown in \ref{fig:chembl2403108-gen-dis}a. This figure reveals the preferred affinities of the generated compounds is from -9.8 kcal/mol to -10.8 kcal/mol, which is also the most popular affinity region of the training samples. Figure \ref{fig:chembl2403108-gen-dis}b illustrates their similarity score distribution to the reference drug Ceritinib.

\begin{figure}[!htb]
	\centering
	\includegraphics[width=0.85\textwidth]{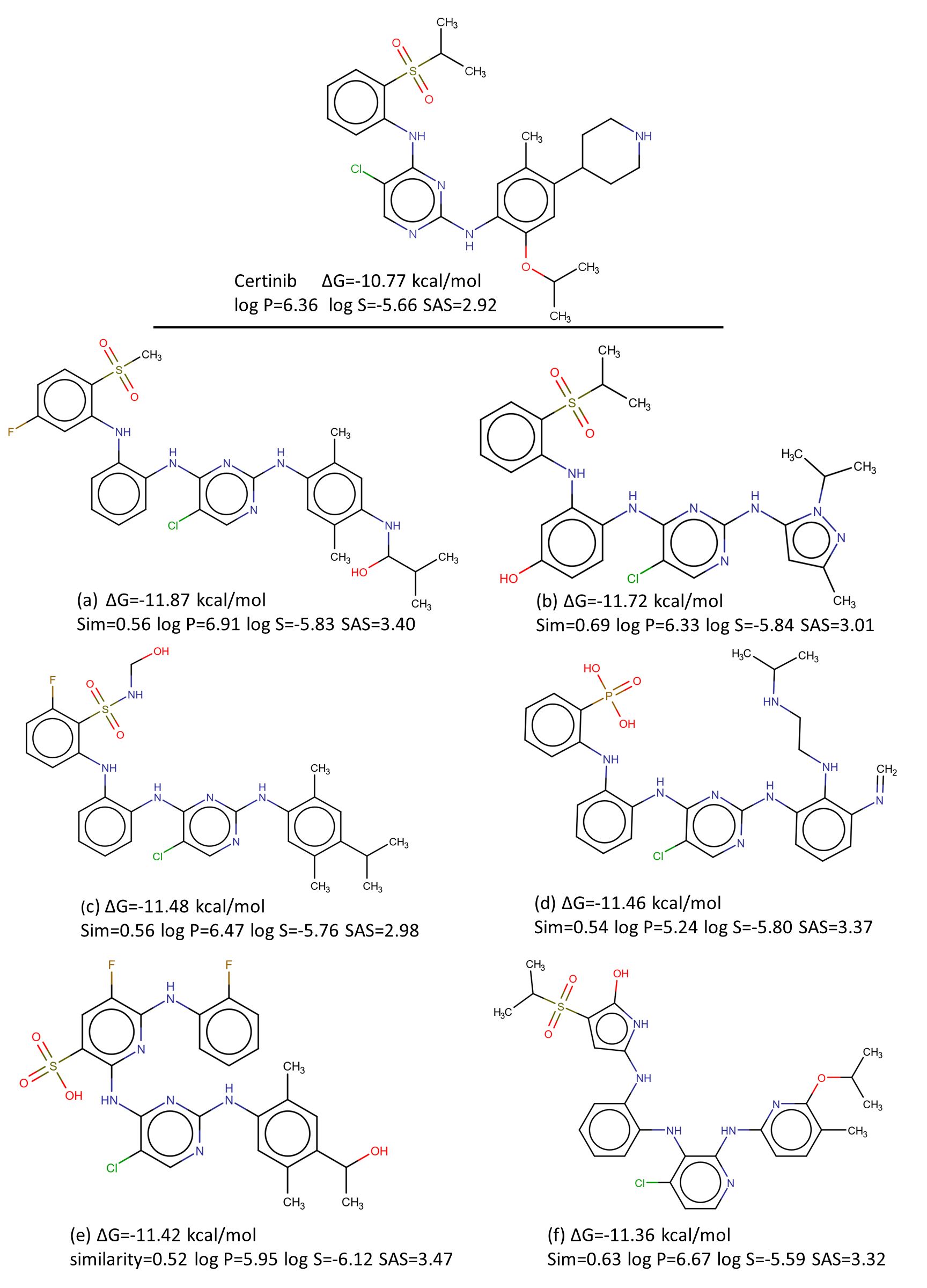}
	\caption{The drug Ceritinib and its top 6 generated molecules. The predicted $\Delta G$s to the target ALK, similarity scores (Sim) to the drug, calculated log P, log S values, and synthetic accessibility scores (SASs) are also present.}
	\label{fig:chembl2403108-top6}
\end{figure}

\paragraph{Top 6 drug candidates.} Ranked by the average predicted binding affinities of these 629 molecules, we select the top 6 drug candidates. Their 2D draws are plotted in Figure \ref{fig:chembl2403108-top6}. The relative errors between their LV-BP and 2DFP-BP predictions are 1.3 \%, 3.1 \%, 0.1 \%, 4.0 \%, 3.5 \%, 3.9 \%, respectively.  It is delighted to see that their average predicted affinities are much higher than that of the reference drug Ceritinib. Moreover, these candidates have similarity scores to the reference drug  from 0.54 to 0.69. 
 They are brand new and do not exist in the SciFinder database. We also calculate their log P values and synthetic accessibility scores (SAS) using RDKit, log S values by Alog PS 2.1 \cite{tetko2005virtual}, and report them in Figure \ref{fig:chembl2403108-top6}; these values are comparable to that of the reference drug and in reasonable ranges.

\paragraph{The interaction and pharmacophore analysis.} One can observe from Figure \ref{fig:chembl2403108-top6}, even though the similarity scores are only between 0.56 and 0.69, these generated compounds share some moieties with the reference drug Ceritinib. This designates these common moieties to possibly involve critical interactions with the binding site of the target.

\begin{figure}[ht!]
	\centering
	\includegraphics[width=0.8\textwidth]{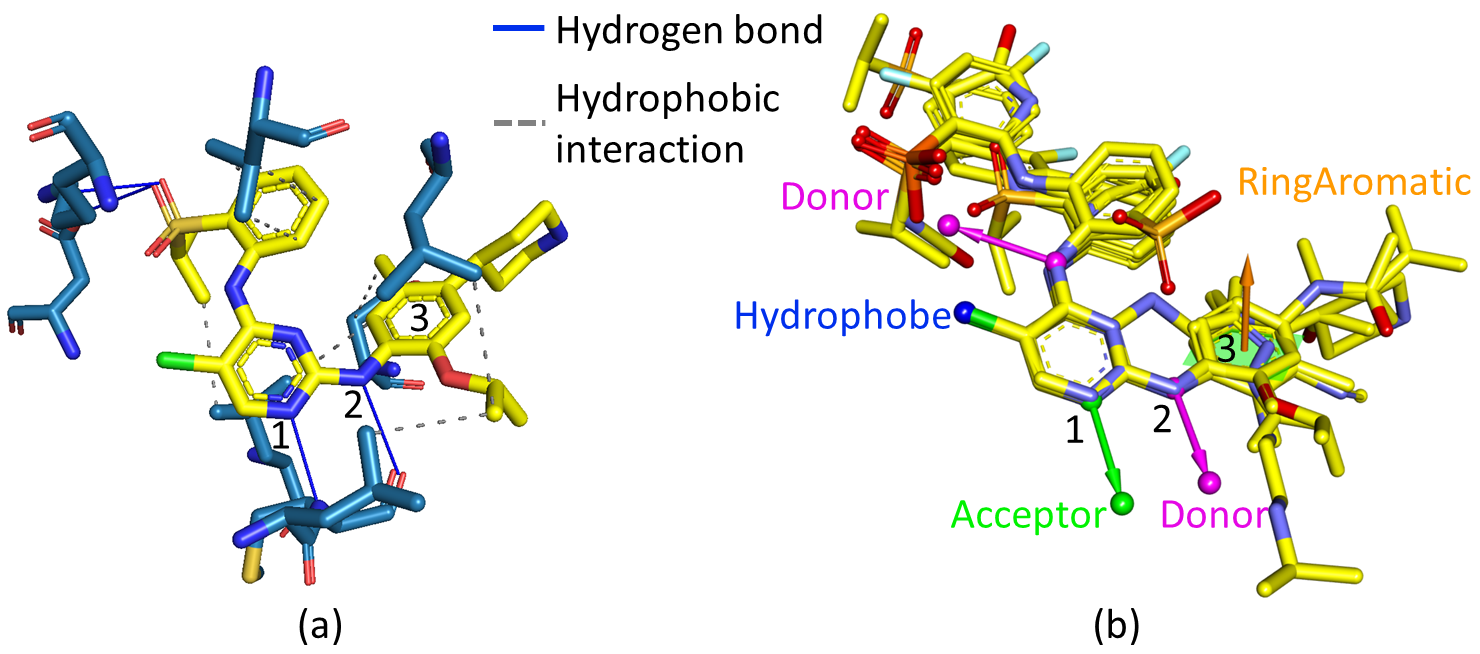}
	\caption{The pharmacophore analysis to the ALK tyrosine kinase's binding site. (a) The interaction plot between the drug Ceritinib and ALK tyrosine kinase from the 3D experimental structure with the PDB ID 4MKC. (b) The 3D alignments of the common pharmacophores obtained from all the active compounds to the target ALK with the 3D experimental structure of the drug Ceritinib, the top 6 generated molecules are also aligned to it.}
	\label{fig:interaction-pharm-chem2403108}
\end{figure}

To verify our generated compounds still contain critical moieties to the binding, from experimental structures, we investigate the drug-target interactions, as well as the common pharmacophores among all the active compounds to the target. Figure \ref{fig:interaction-pharm-chem2403108}a shows the interaction details between the drug Ceritinib and its target in the 3D crystal structure of their complex (PDB ID 4MKC \cite{friboulet2014alk}). It reveals their interactions include, one hydrogen bond with one N atom in the pyrimidine of the drug (marked by 1 in the Figure \ref{fig:interaction-pharm-chem2403108}a) as the acceptor, one hydrogen bond with one N atom in the chain of the drug (marked by 2 in the Figure \ref{fig:interaction-pharm-chem2403108}a) as the donor, and another hydrogen bond with one O atom in the drug as the acceptor, one hydrophobic interaction between one benzene ring (marked by 3 in the Figure \ref{fig:interaction-pharm-chem2403108}a) and the target, as well as other hydrophobic interactions.

The common pharmacophore analysis in Figure \ref{fig:interaction-pharm-chem2403108}b is consistent with the interaction analysis. The N atom in the pyrimidine of the drug (marked by 1 in the Figure \ref{fig:interaction-pharm-chem2403108}a and \ref{fig:interaction-pharm-chem2403108}b) and the N atom in the chain of the drug (marked by 2 in the Figure \ref{fig:interaction-pharm-chem2403108}a and \ref{fig:interaction-pharm-chem2403108}b)) are critical pharmacophores, they plays the roles of an acceptor and a donor of hydrogen bonds, respectively. Another pharmacophore is the benzene ring forming a hydrophobic interaction with the target. The pharmacophore analysis also reveals more potential interaction modes, such as the hydrophobic interaction between the Cl atom and target, and the hydrogen bond with the other N atom in the chain of the drug as its donor.

As illustrated in Figure \ref{fig:interaction-pharm-chem2403108}b, all these critical pharmacophores are retained in our top 6 generated compounds, which strongly supports that these compounds are potential inhibitors to the target.

\subsubsection{The double-target drug: Ribociclib}

\paragraph{The statistics of the drug Ribociclib.} Here, we test the generative power of our GNC on multi-target drugs. In this case study, the multi-property restraints consist of multiple binding affinity criteria and the similarity score to a reference molecule. The drug we test here is Ribociclib (ChEMBL ID: CHEMBL3545110, brand name: Kisqali). It was developed by Novartis and Astex Pharmaceuticals and approved by the FDA in 2017 to treat certain kinds of breast cancers. The monthly cost of Ribociclib treatment is \$10,950 in the US.

\begin{figure}[!htb]
	\centering
	\includegraphics[width=0.8\textwidth]{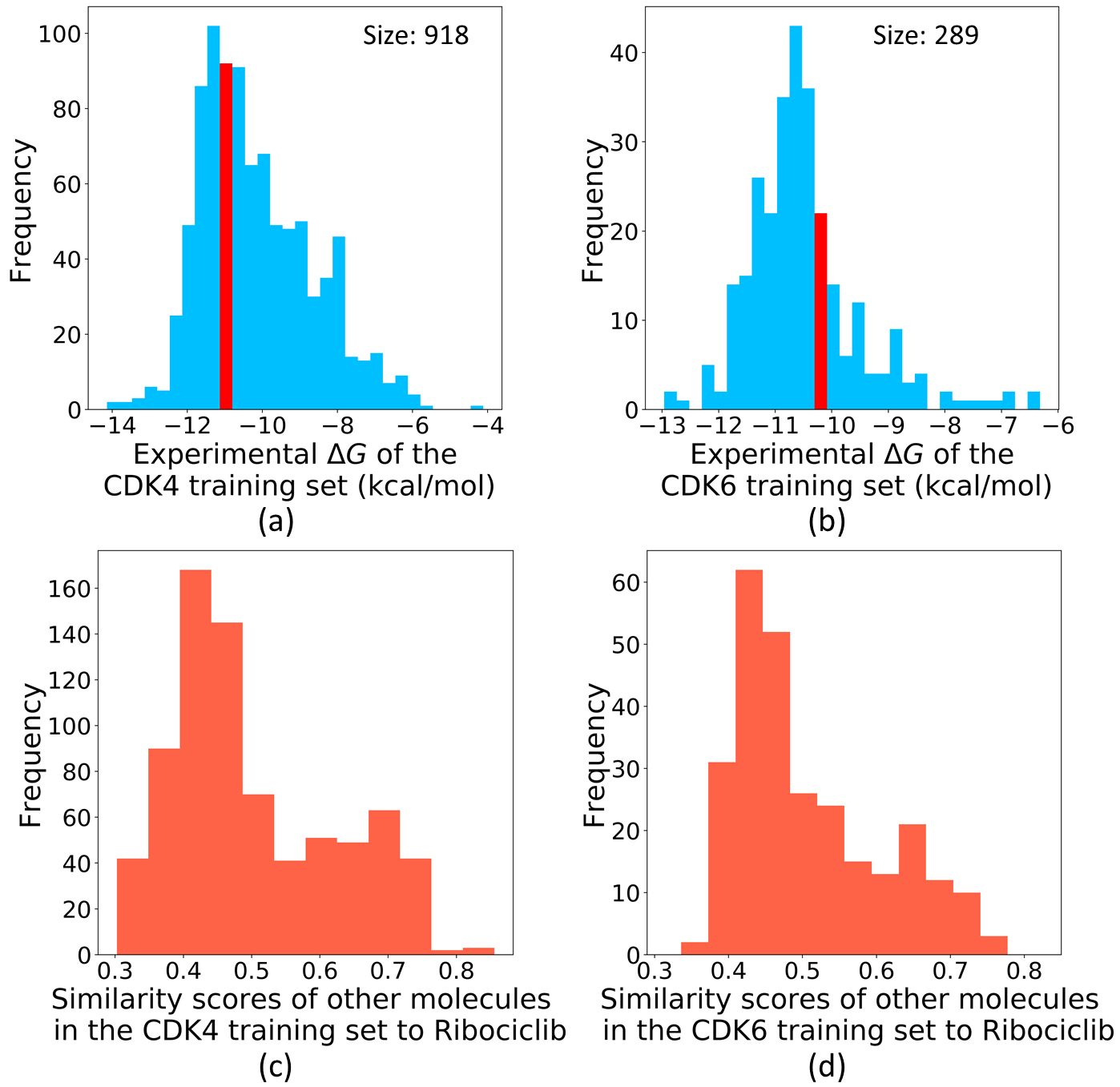}
	\caption{(a) The experimental $\Delta G$ distribution of the training set to the target CDK4 with the interval containing the $\Delta G$ of the drug Ribociclib to CDK4 marked in red. (b) The experimental $\Delta G$ distribution of the CDK6 training set with the interval containing the $\Delta G$ of Ribociclib to CDK4 in red. (c) The similarity score distribution of the other samples in the CDK4 training set to Ribociclib. (d) The similarity score distribution of the other samples in the CDK6 set to Ribociclib.}
	\label{fig:chembl3545110-training-dis}
\end{figure}

Ribociclib inhibits two different targets, namely the Cyclin-dependent kinase 4 (CDK4, CHEMBL ID: CHEMBL331) and Cyclin-dependent kinase 6 (CDK6, CHEMBL ID: CHEMBL2508). In the ChEMBL database, 919 molecules have CDK4 binding data, 289 molecules have CDK6 binding data. After filtered out, 918 and 289 molecules are retained, respectively, providing a small training set to the CDK6. Figure \ref{fig:chembl3545110-training-dis}a and b plot the $\Delta G$ distributions of these two training sets. It reveals, in both the CDK4 and CDK6 sets, hundreds of samples have binding affinities close to or even higher than Ribociclib. Figure \ref{fig:chembl3545110-training-dis}c and d show the similarity score distributions to Ribociclib of the two training sets. Both these sets include more than 200 samples with the similarity scores to the drug over 0.3, more than 60 samples with such scores over 0.6. 

\paragraph{The multitask predictor.}
Since the CDK6 training set is small, it is challenging to train an accurate predictor for this target. However, the two targets, CDK4 and CDK6, are similar due to the calculation via SWISS-MODEL \cite{schwede2003swiss} with the sequence identity being 71.1\%. Therefore, multitask deep learning can enhance reliability.

In our multitask architecture, the binding affinity predictor in our generator offers two outputs, each for one of the two targets, so the predictor is trained by the two training sets simultaneously. As a result, in a 5-fold crossing-validation test, the multitask model significantly improves the performance on the small dataset.

\begin{table}[!htb]
	\centering
	\begin{tabular}{|l|l|l|l|l|}
		\hline
		\rowcolor{Lightgreen}
		\multirow{2}{*}{}    & \multicolumn{2}{l|}{Target CDK4} & \multicolumn{2}{l|}{Target CDK6} \\
		\cline{2-5}
		\rowcolor{Lightgreen}
		& Single task        & Multitask        & Single task         & Multitask        \\ \hline
		LV-BP & 0.791              & 0.804            & 0.524               & 0.811            \\ \hline
		2FP-BP      & 0.824              & 0.836            & 0.485               & 0.779            \\ \hline
	\end{tabular}
	\caption{The $R_p$s of $\Delta G$ predictions from the 5-fold cross-validation tests on the two targets of the drug Ribociclib by the single task and multitask predictors.}
	\label{table:5-fold_ST_MT}
\end{table}

As illustrated in Table \ref{table:5-fold_ST_MT}, the target CDK4 with a 918-molecule training set does not benefit so much from the multitask. However, to the target CDK6 only with a 289-molecule training set, the improvement is dramatic: the $R_p$s rise from 0.524 to 0.811 by the LV-BP and from 0.485 to 0.779 by the 2FP-BP. These results demonstrate the efficiency of the multitask architecture.

\paragraph{The generation of new druglike molecules.}
To design alternative Ribociclib drugs in broader chemical space, we set the similarity score restraints to Ribociclib from 0.30 to 0.90, with an increment of 0.025. The binding affinities to the target CDK4 are aimed at the interval between -10.80 kcal/mol and -12.00 kcal/mol with an increment of -0.2 kcal/mol; this interval covers the $\Delta G$s of Ribociclib as well as lots of other training samples. For the same reason, the objectives of the binding affinities to the target CDK6 are set from -10.2 kcal/mol to -11.0 kcal/mol with an increment of -0.2 kcal/mol. Totally, following this scheme, we create 1080 novel molecules.

\paragraph{The reevaluation by 2DFP-BP.}
\begin{figure}[!htb]
	\centering
	\includegraphics[width=0.9\textwidth]{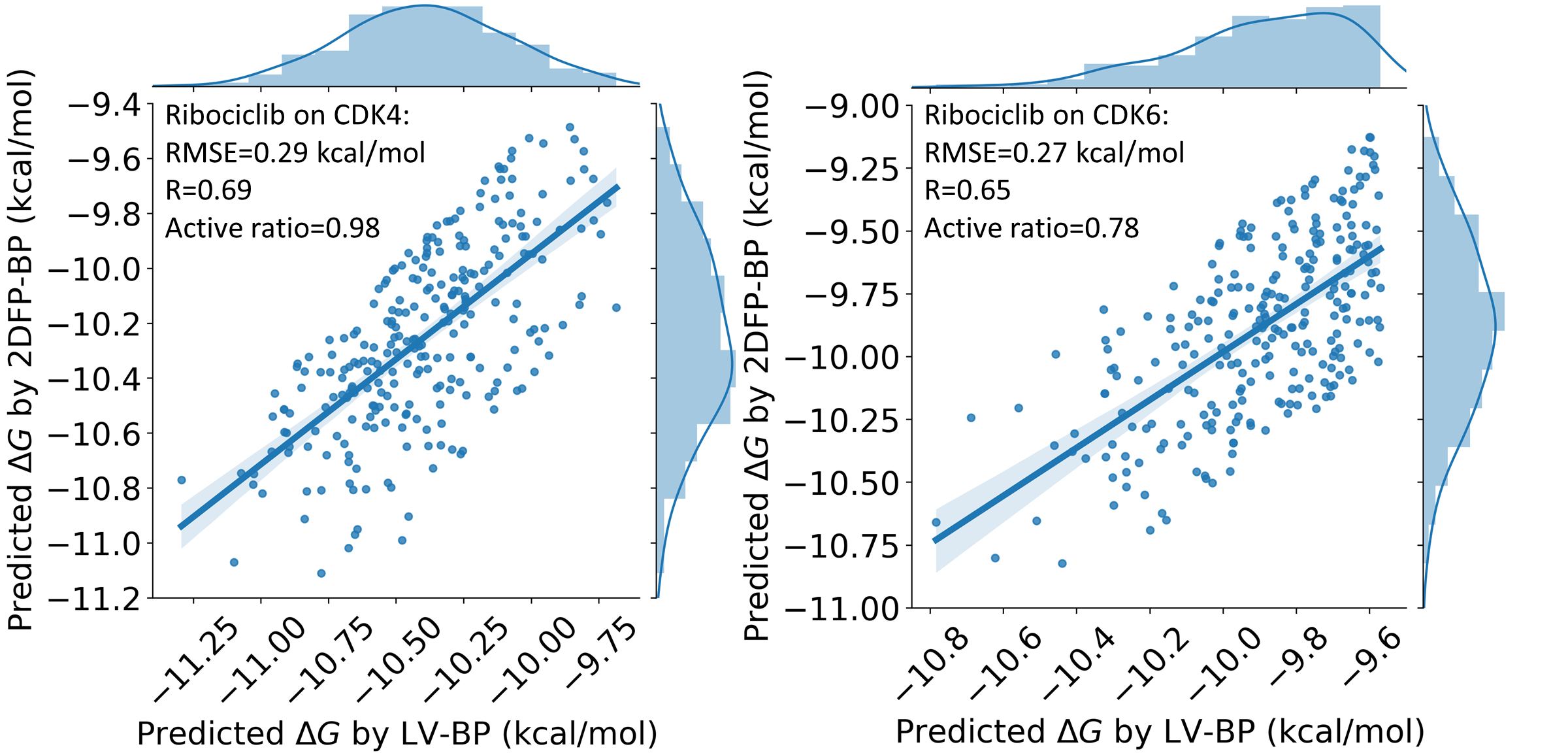}
	\caption{The correlations between the LV-BP and 2DFP-BP predicted $\Delta G$s of the generated molecules to the targets CDK4 and CDK6, the ones having high relative errors between the two predictions (>5 \%) are already eliminated.}
	\label{fig:corr-2d-dl-two-targets}
\end{figure}

The predicted bind affinities of these 1080 generated compounds are reevaluated by the 2DFP-BP model incorporated with the multitask DNN. The parameters of this architecture are introduced in Section \ref{sec:param-multi}.  
After excluding the ones with high discrepancies between the LV-BP and 2DFP-BP predictions, 271 molecules are left.

Figure \ref{fig:corr-2d-dl-two-targets} depicts the correlation plots between their LV-BP and 2DFP-BP predicted $\Delta G$s to the two targets. The correlations of the $\Delta G$s to both the two targets are promising. Specifically, the $\Delta G$s to the target CDK4 can achieve an RMSE of 0.29 kcal/mol, a $R_p$ of 0.69, and an active ratio as high as 0.98; the $\Delta G$s to the target CDK6 also have an RMSE of 0.27 kcal/mol, a $R_p$ of 0.65 and an active ratio of 0.78.

\begin{figure}[!htb]
	\centering
	\includegraphics[width=0.8\textwidth]{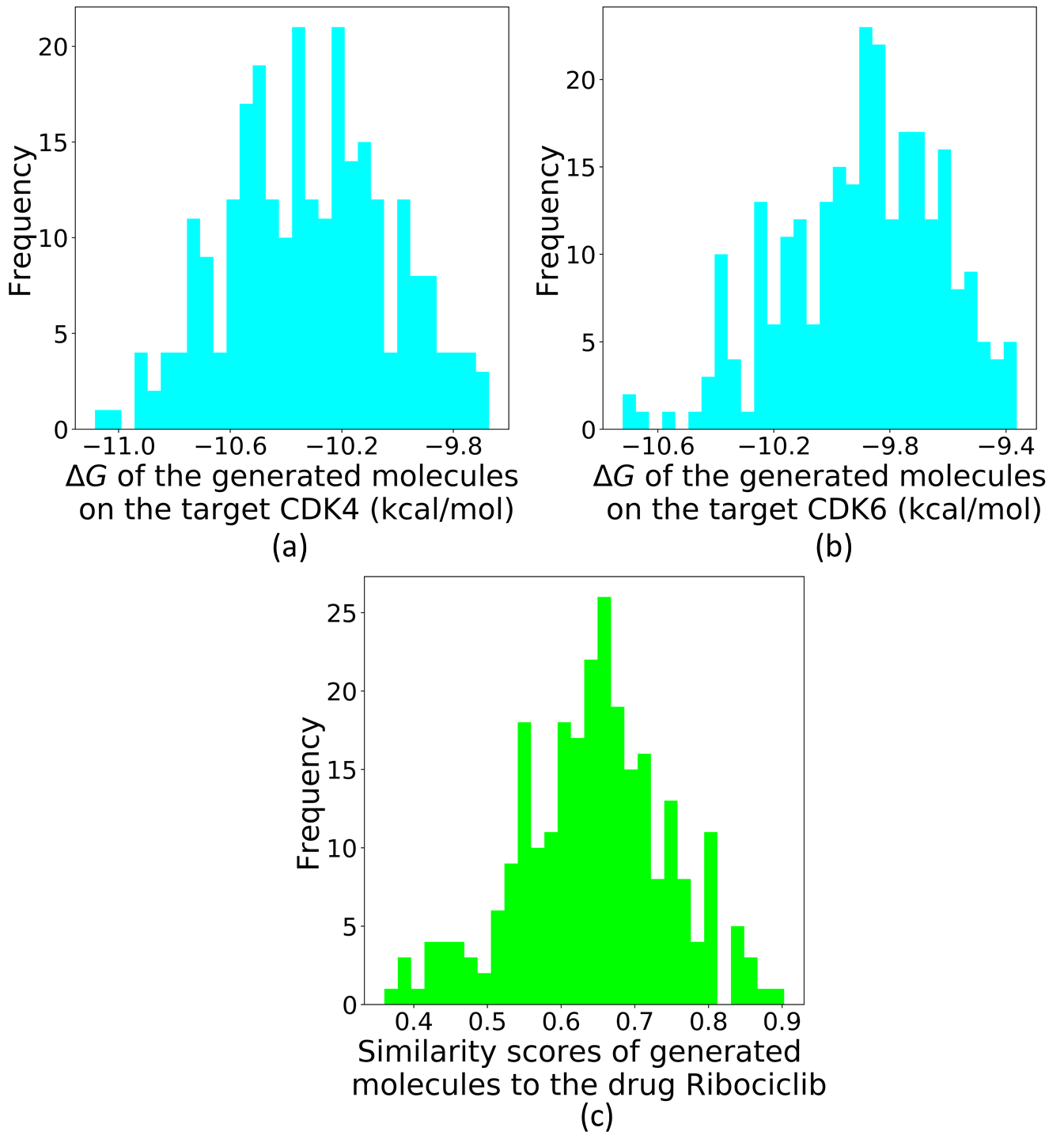}
	\caption{The average predicted $\Delta G$ to the targets CDK4 and CDK6 of the generated molecules and their similarity scores to the reference drug Ribociclib: (a, b) The distributions of the averages of the LV-BP and 2DFP-BP predicted $\Delta G$s to the two targets, respectively. (c) The similarity score distribution to the reference drug Ribociclib.}
	\label{fig:chembl3545110-gen-dis}
\end{figure}

Their LV-BP and 2DFP-BP predicted binding affinities to the two targets are also averaged, and the distributions are shown in Figure \ref{fig:chembl3545110-gen-dis}a,b. Figure \ref{fig:chembl3545110-gen-dis}c illustrates their similarity score distribution to the reference drug Ribociclib.

\begin{figure}[!htb]
	\centering
	\includegraphics[width=0.7\textwidth]{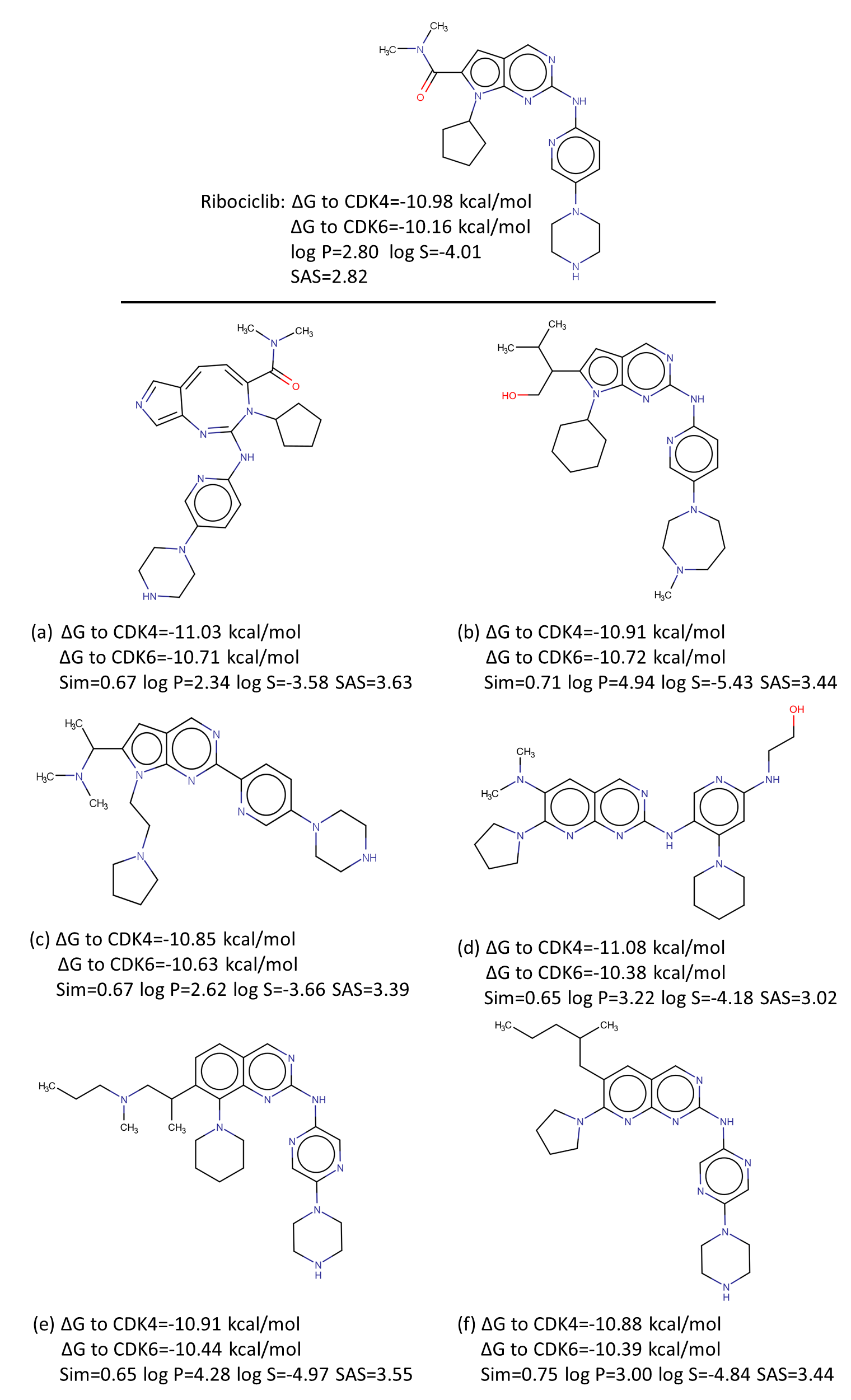}
	\caption{The drug Ribociclib and its top 6 generated molecules. The predicted $\Delta G$s to the targets CDK4 and CDK6, similarity scores (Sim) to the drug, calculated log P, log S values, and synthetic accessibility scores (SASs) are also reported.}
	\label{fig:chembl3545110-top6}
\end{figure}

\paragraph{Top 6 drug candidates.}

According to the average predicted binding affinities of these 271 molecules, we select the top 6 drug candidates. Figure \ref{fig:chembl3545110-top6} represents their 2D plots. To the target CDK4,  the relative errors between their LV-BP and 2DFP-BP predictions are 4.6 \%, 3.0 \%, 3.2 \%, 0.2 \%, 1.6 \% and 0.7 \%, respectively; to CDK6, the relative errors between the two predictions are 1.7 \%, 1.2 \%, 3.7 \%, 3.4 \%, 4.8 \% and 1.7 \%, respectively. Most of them have better binding affinities than the reference drug Ribociclib. Such as, to CDK4, the affinities of the first and fourth compounds are predicted to be higher than that of Ribociclib, the other three have similar ones with Ribociclib; to CDK6, all the top 6 candidates have better binding affinities. Moreover, their similarity scores to the reference drug are between 0.65 and 0.75. They are not in the SciFinder database, which means these six candidates are novel. The log P, log S values and synthetic accessibility scores (SASs) calculated by RdKit and Alog PS are also showed in the figure; their log P, log S values and SASs are comparable to that of the reference drug.

\begin{figure}[!htb]
	\centering
	\includegraphics[width=1.0\textwidth]{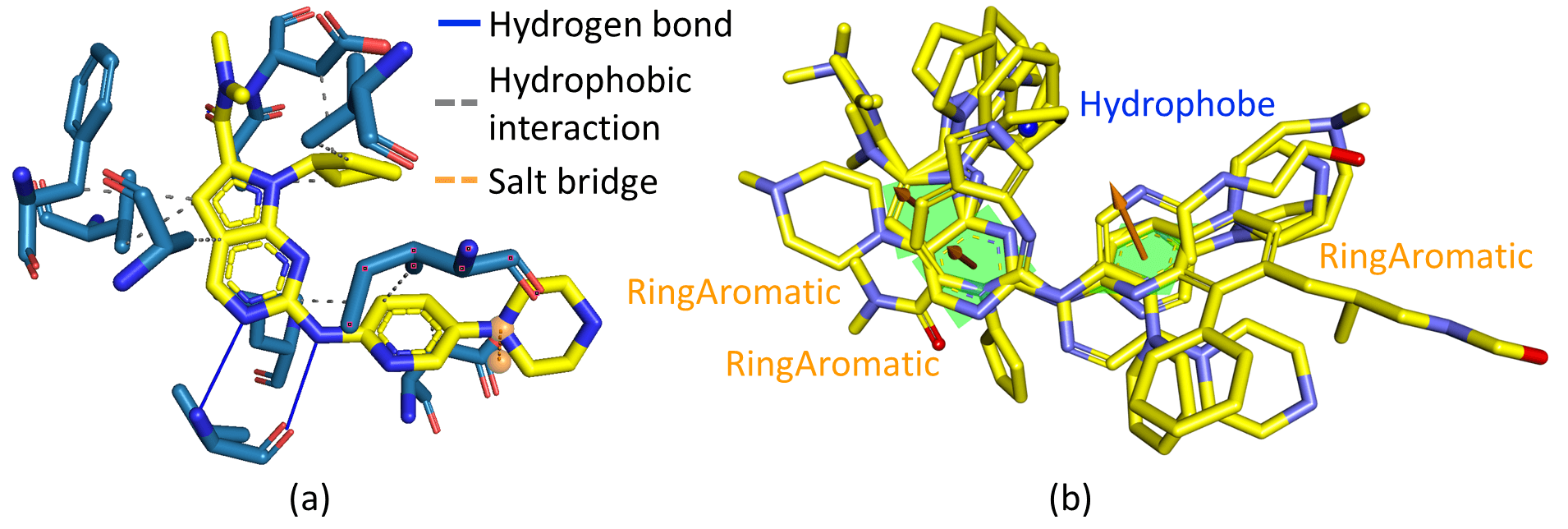}
	\caption{(a) The illustration of the interactions between the drug Ribociclib and the target CDK6 extracted from the experimental structure (PDB ID 5L2T). (b) The 3D alignment of the common pharmacophores obtained from all the active compounds to the target CDK6 with the 3D experimental structure of the drug Ribociclib, the top 6 generated molecules are also aligned to it.}
	\label{fig:interaction-pharm-3545110-target2}
\end{figure}

\paragraph{The interaction and pharmacophore analysis.}
Figure \ref{fig:interaction-pharm-3545110-target2}(a) shows the interaction details between the drug Ribociclib and the target CDK6 in the 3D crystal structure of their complex (PDB ID 5L2T \cite{chen2016spectrum}). It indicates the main interactions are hydrogen bonds and hydrophobic interactions. The pharmacophore analysis in Figure \ref{fig:interaction-pharm-3545110-target2}(b) reveals that the critical pharmacophores are the pyrrolopyrimidine, pyridine, and cyclohexane, which can form hydrogen bonds and hydrophobic interactions with the binding site. 

Among the top 6 candidates, except the first one, the others contain all these critical pharmacophores; the first one has most of them. This suggests all the six compounds are potential inhibitors to the targets.

\begin{figure}[!htb]
	\centering
	\includegraphics[width=0.8\textwidth]{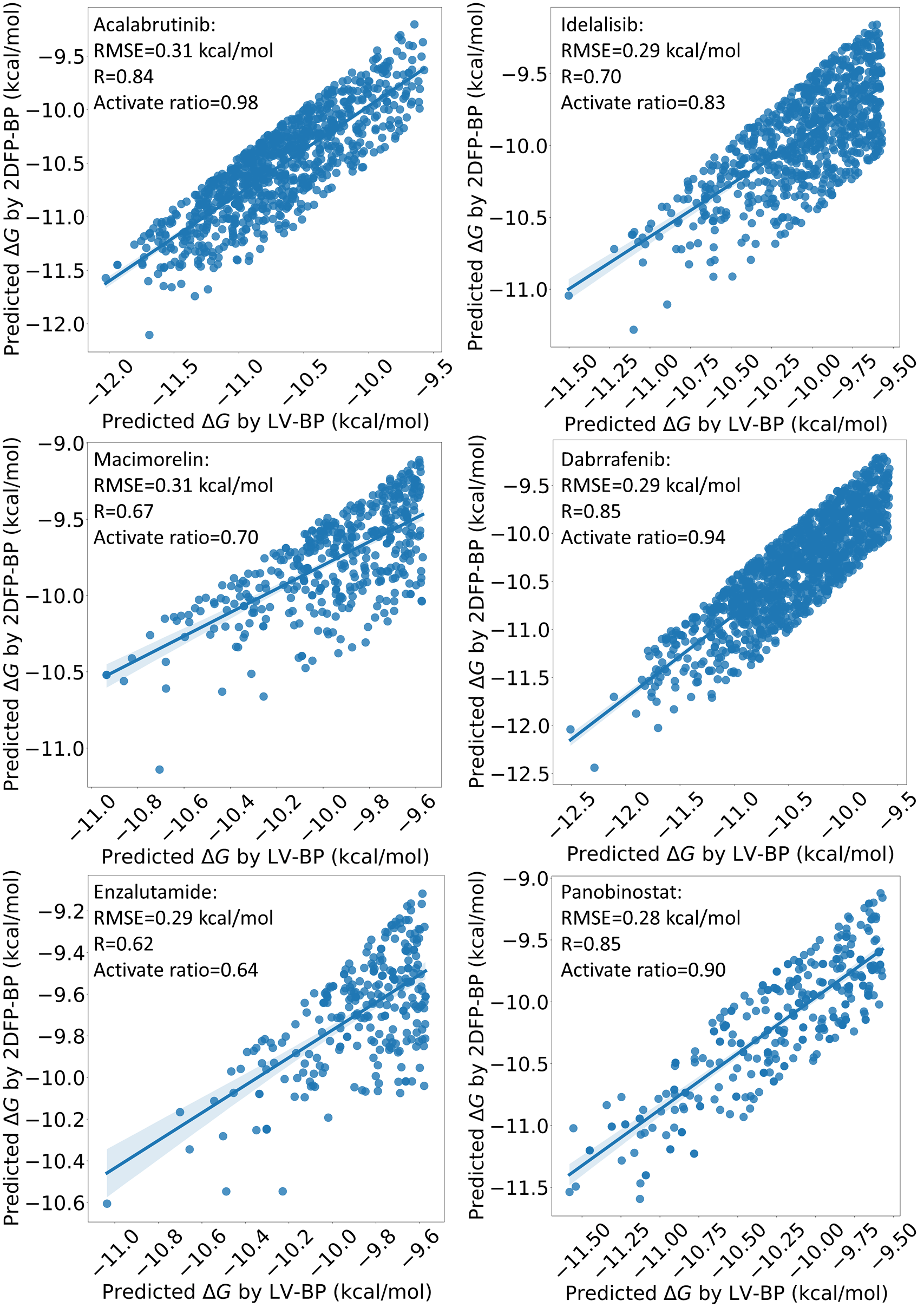}
	\caption{The correlation plots between the LV-BP and 2DFP-BP predicted $\Delta G$s of the generated molecules to the six drug targets, the ones with high relative errors between the two predictions are already eliminated.}
	\label{fig:corr-other-drugs}
\end{figure}

\subsubsection{The tests on the other single-target drugs}

We also apply our GNC to the rest six drugs listed in Table \ref{table:eight-drug-inf} and design novel drug candidates. The similarity score restraints are from 0.30 to 0.90, with an increment of 0.025. The $\Delta G$ objective ranges are chosen to contain the $\Delta G$s of the drugs as well as lots of other training samples. We also only collect the generated molecules with the relative errors between the LV-BP and 2DFP-BP predicted $\Delta G$s below 5 \%.

\begin{figure}[!htb]
	\centering
	\includegraphics[width=0.9\textwidth]{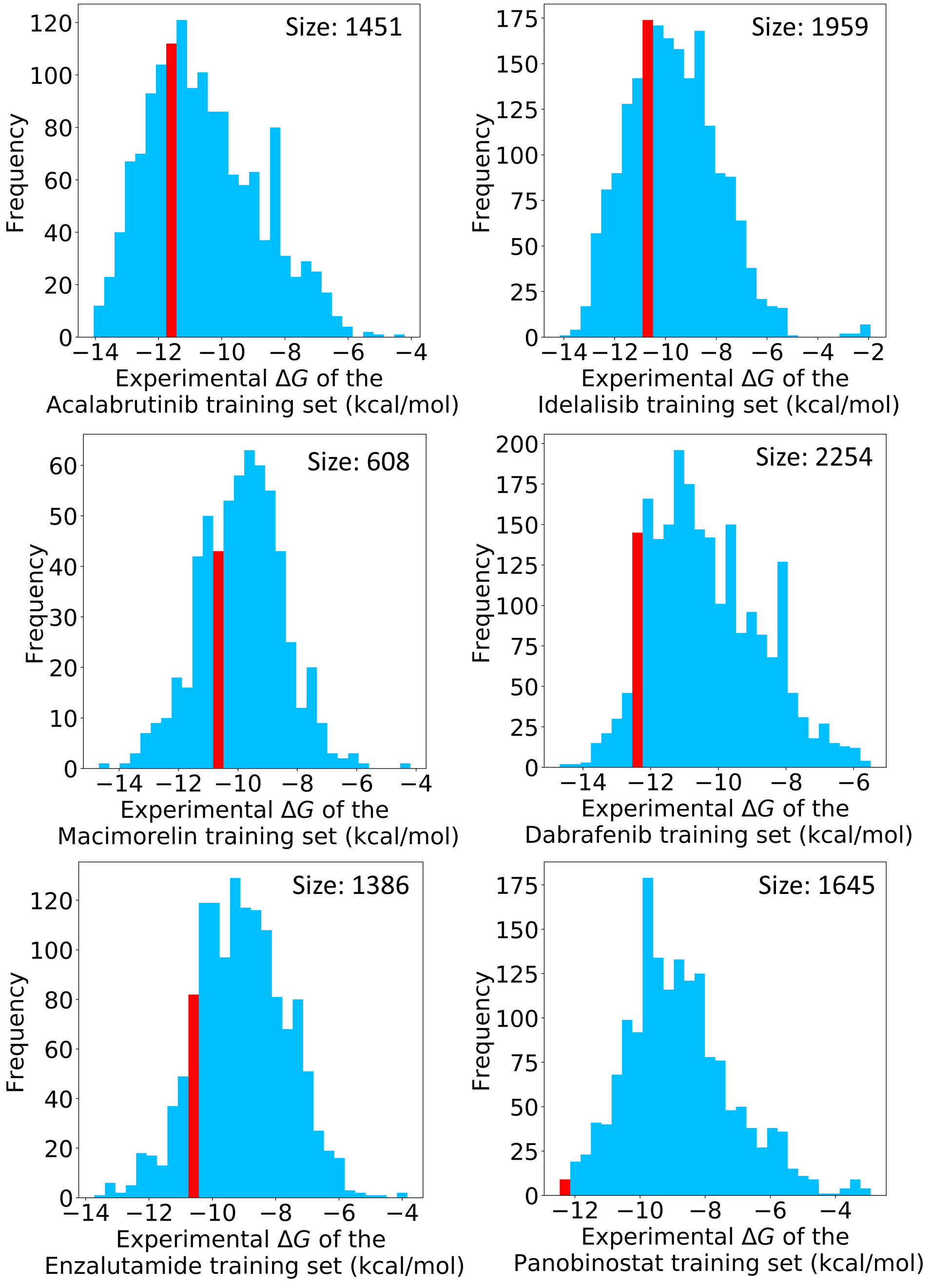}
	\caption{The sizes and experimental $\Delta G$ distributions of the training sets to the six drug targets. The red bars indicate the intervals containing the $\Delta G$s of the six reference drugs.}
	\label{fig:detg-dis-others}
\end{figure}

\begin{figure}[!htb]
	\centering
	\includegraphics[width=0.9\textwidth]{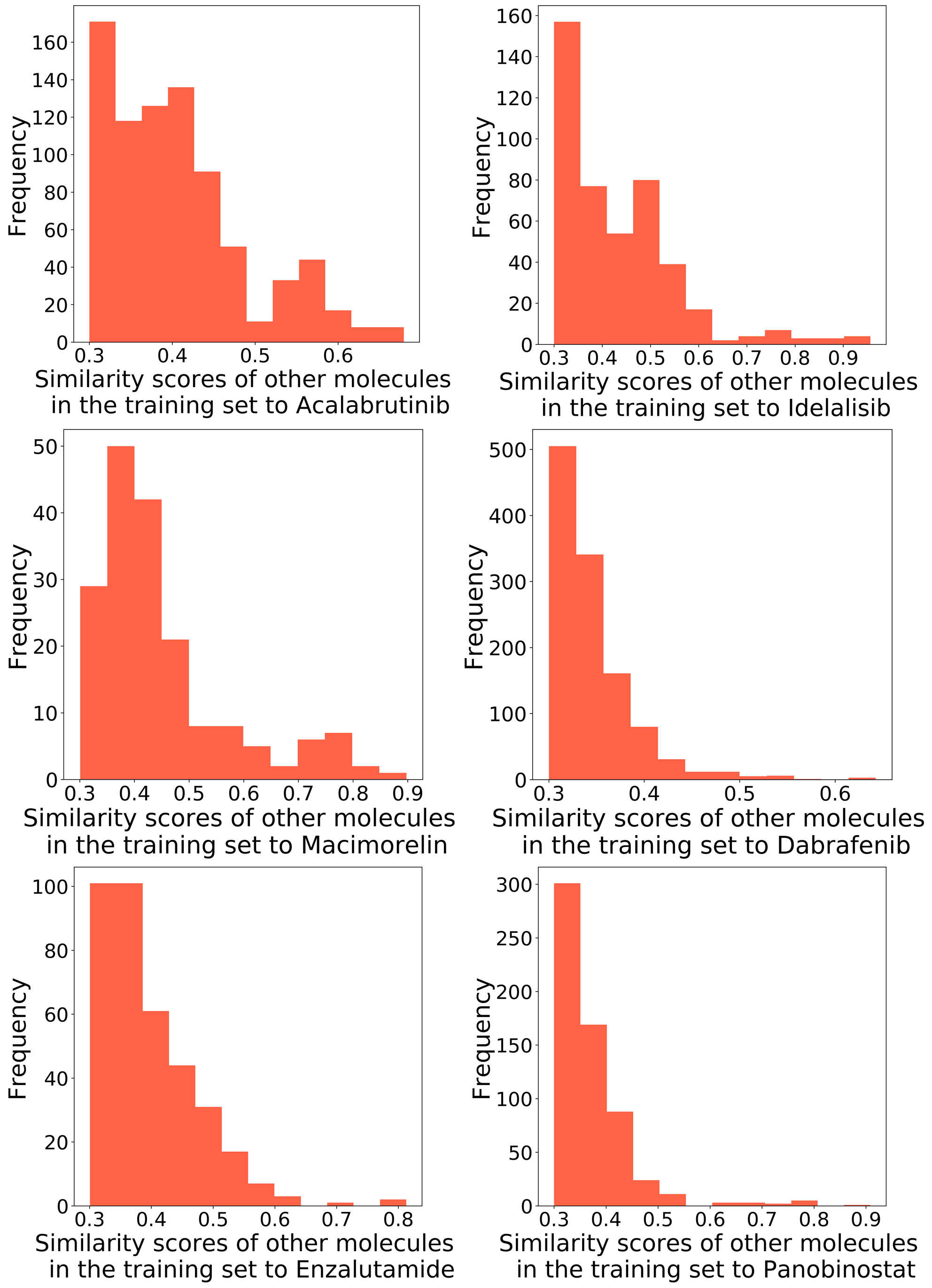}
	\caption{ The similarity score distributions of other molecules in the six training sets to the reference drugs.}
	\label{fig:sim-dis-others}
\end{figure}

Figure \ref{fig:corr-other-drugs} indicates, since the ones with high discrepancies between the two predictions are eliminated, our generated compounds to all these six drug targets have promising correlations. 

Now, we focus on whether highly active drug candidates are created or not. This relies on the binding affinity and similarity score distributions of the training sets. Figure \ref{fig:detg-dis-others} provides the sizes and the $\Delta G$ distributions of these six target-specified training sets; the red bars indicate the intervals containing the $\Delta G$s of the reference drugs. Figure \ref{fig:sim-dis-others} plots their similarity score distributions.

Figures \ref{fig:detg-dis-others}a shows that the training set of the drug Acalabrutinib contains 1459 compounds. Among them, more than 400 compounds have higher binding affinities than that of the drug, which is beneficial to generate compounds more potent than the drug. Also, Figure \ref{fig:sim-dis-others}a reveals that in the training set, over 500 molecules have the similarity scores to the drug over 0.3, which is quite helpful to generate novel compounds with similarity scores also in this range. These promising statistical information helps to explain why our GNC can create as many as 879 new compounds for this drug target with high confidence, and 17 of them have higher or similar binding affinities with that of Acalabrutinib; the best $\Delta G$ is -11.90 kcal/mol (see Figures S1 and S3).

As revealed by \ref{fig:detg-dis-others}b, the training set of the drug Idelalisib is larger than that of Acalabrutinib; and over 500 molecules in this set have higher binding affinities than that of Idelalisib. As a result, 73 of the 794 generated compounds are more potent than Idelalisib, with the best $\Delta G$ being -11.27 kcal/mol (see Figures S1 and S4).

Figure \ref{fig:detg-dis-others}c exhibits that the dataset of Macimorelin is quite small. However, the correlation of the generated molecules is still satisfactory, and 13 generated molecules have similar $\Delta G$s with that of the drug (see Figure S1). The reason is, its training set contains 205 compounds with higher binding affinities than that of the drug, and also 182 compounds with the similarity scores to the drug over 0.3.

Figures \ref{fig:detg-dis-others}d and e depict the datasets of the drugs Dabrafenib and Enzalutamide. The affinity values of these two drugs are close to the upper boundaries of their training sets' affinity domains. In other words, few molecules in their training sets are more active than the drugs. Since the interpolation nature of machine learning models tends to provide reliable predictions inside the populated range, it is tough to generate compounds more potent than these drugs. Although their training set sizes and similarity scores are favorable, the numbers of generated compounds with better binding affinities than that of the drugs are only 2 and 4, respectively (see Figures S1, S6 and S7).

We perform our last experiment on the drug Panobinostat. Panobinostat is an extreme example: as illustrated in Figure \ref{fig:detg-dis-others}f, its binding affinity is the highest in its training set. Therefore, although our GNC model generates 319 molecules, the top active one among them only reaches a $\Delta G$ of -11.56 kcal/mol, which is far less potent than the drug itself ($\Delta G$=-12.46 kcal/mol).

In the supporting information, Figures S3 to S8 provide the top six generated alternative compounds for the drugs Acalabrutinib, Idelalisib, Macimorelin, Dabrafenib, Enzalutamide, and Panobinostat, as well as their predicted $\Delta G$s.

\section{Discussions}\label{sec:discussions}

With the availability of deep learning technologies, more and more {\it in silico} molecule generation models have been proposed. These models can be classified into three categories: randomized output, controlled output, and optimized output \cite{grow2019generative}. One of the challenges is how to generate new molecules with desired chemical properties, especially drug-like molecules. Another challenge is how to improve the utility of {\it in silico} molecule generation without direct experimental validations.  
To address these challenges, we propose a new GNC to generate drug-like molecules based on multi-property optimizations via gradient descent. 

\subsection{The essential circumstances for reliable and desired molecule generation}
Based on our experiments, there are two essential circumstances to generate molecules with reliable and desired predicted chemical properties:

\begin{enumerate}
\item An objective property value should always be in a region with lots of training samples. Based on the nature of machine learning methods, the predictor can be built with high accuracy at an objective value if lots of training samples around it. It can be seen from Section \ref{sec:BACE1-reliablity-test} that when a binding affinity objective is in the middle of the training set's distribution, the latent space generator can always generate novel molecules with reliable predictions confirmed by the 2DFP-BP. However, when an objective is close to or even at the edge of the training-set distribution, one may still generate many novel compounds, but the predictions to them are quite risky, such as high discrepancies between the LV-BP and 2DFP-BP predictions.  
\item Generated compounds should have some high similarity scores to some existing molecules in the training set. If some molecules (not necessarily reference molecules) in the training set are similar to generated compounds, then the predictions are reliable and can be verified by the 2DFP-BP. 
 \end{enumerate}

\subsection{The necessity of both similarity and property value restraints}
The two points above also explain why both similarity score and property value restraints should be included in our generator.

The goal of property restraint is two-fold. First, it restricts property values to desired ones. Additionally, it can be used to achieve high reliability. As discussed above, if the property value is limited to a populated region of the training-set distribution, the resulting generated molecules will most likely have reliable predictions.

Similarity restraint is also to ensure prediction accuracy. High similarity scores to existing molecules make predictions more accurate and reliable. Moreover, in a regular drug discovery procedure for a given target, one typically starts from some lead compounds or even current drugs and then carries out lead optimizations to make candidates more "drug-like", e.g.,  higher activity and lower side effects \cite{hughes2011principles}. Therefore, similarly, in a drug-specified generative model, the similarity to a reference compound or drug must be controlled to guarantee that new drug-like compounds still bind to the target. For example, with similarity restraint, generated molecules share pharmacophores with the reference drug.

\subsection{Multiple property restraints}

Drug design is sophisticated. To develop a drug, plenty of properties must be carefully studied, such as binding affinity, toxicity, partition coefficient (log P), aqueous solubility (log S), off-target effect. The failure in any one of these properties can prevent drug candidates from the market. In other words, drug design is a multi-property optimization process.

Technically, our generator can handle this multi-property optimization. In our framework, the restraint to each property is realized by one term in the loss function.  Therefore, multi-property optimization can be satisfied simultaneously in our GNC. In this work, multiple property restraints are tested on one drug with two targets (Ribociclib). It turns out, the generated new candidates have ideal binding affinities to the two targets simultaneously; this means, our model can work on the multiple property restraints. Multiple properties can also be specified as toxicity, log P, log S, etc. To avoid side effects, one can also control drug candidates to have a high binding affinity to one target but low binding affinities to other targets.

\section{Conclusion} \label{sec:conclusion}
Searching alternative drugs is important for improving the quality of existing drugs and making new drugs cheaply available to low-income people.
In this work, we develop a new generative network complex (GNC)  for automated generation of drug-like molecules based on the multi-property optimization via gradient descent in the latent space. In this new GNC, multiple chemical properties, particularly binding affinity and similarity score, are optimized to generate new molecules with desired chemical and drug properties. 
To  ensure the prediction reliability of these new compounds, we reevaluate them by independent 2D-fingerprint based predictors. Molecules without consistent predictions from the latent-vector model and the 2D-fingerprint model are not accepted. After the consistent check, hundreds of potential potent drug candidates are reported. Performed on a supercomputer, the generation process  from one seed to a large number of new molecules takes less than 10 minutes. Therefore, our GNC is an efficient new paradigm for discovering new drug candidates.  To demonstrate the utility of the present GNC, we first test its reliability on the BACE1 target and then, further apply this model to  generate thousands of new alternative drug candidates for a few market existing drugs, namely, Ceritinib, Ribociclib, Acalabrutinib, Idelalisib, Dabrafenib, Macimorelin, Enzalutamide, and Panobinostat. 
 
We also discuss the keys to generate drug-like candidates with reliable predictions. First, an objective property value should be in a populated region of the training-set distribution. Second, generated molecules need to have good similarity scores to some existing compounds in the training set.

\vspace{1cm}
\section*{Supplementary materials}
Figures S1 to S8 are the $\Delta G$ and similarity score distributions of the generated drug-like molecules for the drugs Acalabrutinib, Idelalisib, Dabrafenib, Macimorelin, Enzalutamide, and Panobinostat, as well as the 2D plots and predicted $\Delta G$s of the top 6 among them.

\vspace{1cm}
 \section*{Acknowledgments}
This work was supported in part by  NSF Grants DMS-1721024,  DMS-1761320, and IIS1900473,  NIH grant  GM126189, and Michigan Economic Development Corporation. DDN and GWW are also funded by Bristol-Myers Squibb and  Pfizer.

\clearpage
\vspace{1cm}
%

\end{document}